\documentclass[aps,prl,amsmath,amssymb,twocolumn,groupedaddress,longbibliography,floats,final,postprint,superscriptaddress]{revtex4-1}
\usepackage{amsmath,amssymb,amsfonts,float,graphics,epsfig,epstopdf,color,verbatim,tabularx,bm,multirow,appendix}
\usepackage{graphicx}
\usepackage{color}
\definecolor{LinkColor}{RGB}{199,21,133}
\usepackage[colorlinks=true,linkcolor=blue,anchorcolor=blue,citecolor=blue,urlcolor=blue]{hyperref}
\usepackage[polutonikogreek,english]{babel}

\usepackage{array}
\newcolumntype{P}[1]{>{\centering\arraybackslash}p{#1}}

\usepackage{hyperref}
\usepackage{epstopdf}
\usepackage{type1cm}
\usepackage{multirow}
\usepackage{times}
\usepackage{algpseudocode}
\usepackage{algorithm}

\def\scrh{\mathcal{H}}
\def\scrj{\mathcal{J}}
\def\scro{\mathcal{O}}
\def\scrh{\mathcal{H}}

\begin{document}
\title{Extraordinary-log Universality of Critical Phenomena in Plane Defects}
\author{Yanan Sun}
\thanks{Y.S. and M.H. contributed equally to this work.}
\author{Minghui Hu}
\thanks{Y.S. and M.H. contributed equally to this work.}
\affiliation{Department of Physics, Anhui Key Laboratory of Optoelectric Materials Science and Technology, Key Laboratory of Functional Molecular Solids, Ministry of Education, Anhui Normal University, Wuhu, Anhui 241000, China}
\author{Youjin Deng}
\email{yjdeng@ustc.edu.cn}
\affiliation{National Laboratory for Physical Sciences at Microscale, University of Science and Technology of China, Hefei, Anhui 230026, China}
\affiliation{Department of Modern Physics, University of Science and Technology of China, Hefei, Anhui 230026, China}
\author{Jian-Ping Lv}
\email{jplv2014@ahnu.edu.cn}
\affiliation{Department of Physics, Anhui Key Laboratory of Optoelectric Materials Science and Technology, Key Laboratory of Functional Molecular Solids, Ministry of Education, Anhui Normal University, Wuhu, Anhui 241000, China}
\begin{abstract}
The recent discovery of the extraordinary-log (E-Log) criticality is a celebrated achievement in modern critical theory and calls for generalization. Using large-scale Monte Carlo simulations, we study the critical phenomena of plane defects in three- and four-dimensional O($n$) critical systems. In three dimensions, we provide the first numerical proof for the E-Log criticality of plane defects. In particular, for $n=2$, the critical exponent $\hat{q}$ of two-point correlation and the renormalization-group parameter $\alpha$ of helicity modulus conform to the scaling relation $\hat{q}=(n-1)/(2 \pi \alpha)$, whereas the results for $n \geq 3$ violate this scaling relation. In four dimensions, it is strikingly found that the E-Log criticality also emerges in the plane defect. These findings have numerous potential realizations and would boost the ongoing advancement of conformal field theory.
\end{abstract}
\date{\today}
\maketitle

\noindent{\textcolor{blue}{\it Introduction.}---}
In the standard scenario of critical phenomena, the two-point correlation asymptotically decays as $g(r) \sim r^{-(d-2+\eta)}$ with spatial distance $r$, where $d$ and $\eta$ are spatial and anomalous dimensions, respectively~\cite{Stanley1987}.

Recently, a very unusual type of critical phenomena was unveiled in the context of surface critical behavior (SCB)~\cite{metlitski2020boundary}. Consider the O($n$) model of pairwise-interacting unit-vector spins. The cases $n=1$, $2$ and $3$ respectively correspond to the Ising, $XY$ and Heisenberg models. In two dimensions, the Mermin-Wagner theorem prohibits spontaneous symmetry breaking for any finite temperature $T>0$ with $n \geq 2$. Specifically, as $T$ decreases, the $XY$ model enters a quasi-long-range ordered phase via the Berezinskii-Kosterlitz-Thouless (BKT) transition~\cite{berezinskii1971destruction,kosterlitz1972long,kosterlitz1973ordering,kosterlitz1974critical}, while the system with $n>2$ remains disordered for all $T>0$. However, the open surfaces of the critical three-dimensional (3D) O($n$) models, with $n=2$, $3$ and $4$, were observed to undergo a so-called special phase transition and enter the extraordinary phase, as the surface coupling strength is enhanced~\cite{deng2005surface,deng2006bulk}. The nature of the extraordinary phase remained a long-standing puzzle until a recent renormalization-group study~\cite{metlitski2020boundary}, which revealed the extraordinary-log (E-Log) critical phase for $2 \leq n < n_c$, with $n_c$ an upper bound.

In the E-Log critical phase, the correlation function decays as a power law of distance logarithm, $g(r) \sim ({\rm ln}r)^{-(\hat{q}+1)}$, with $\hat{q}+1$ a critical exponent~\cite{metlitski2020boundary}, which is extremely slowly in comparison with $r^{-(d-2+\eta)}$ for the standard scenario. Moreover, for a finite 3D lattice of side length $L$, $g(r,L)$ was numerically observed to exhibit a two-distance scaling behavior as $g(r,L) \sim c_1 ({\rm ln}r)^{-(\hat{q}+1)}+ c_2 ({\rm ln}L)^{-\hat{q}}$, with $c_1$ and $c_2$ being constants~\cite{Hu2021}. In other words, the correlation function decays algebraically with the distance logarithm and then enters into a $L$-dependent plateau, of which the height decreases algebraically with the side-length logarithm (with exponent $\hat{q}$). On the other hand, the helicity modulus $\Upsilon$, characterizing the response against a twist in boundary conditions~\cite{fisher1973}, scales as $\Upsilon L \sim 2\alpha ({\rm ln}L)$, with $\alpha$ a renormalization-group parameter. A scaling relation reads~\cite{metlitski2020boundary,Hu2021}
\begin{equation}\label{SR_qalpha}
\hat{q}=\frac{n-1}{2 \pi \alpha},
\end{equation}
while the original formula means $\hat{q}+1=\frac{n-1}{2 \pi \alpha}$~\cite{metlitski2020boundary}. The E-Log criticality and scaling relation~(\ref{SR_qalpha}) were first verified at $n=3$~\cite{toldin2020boundary}. For the $XY$ model, the universality of $\hat{q}$ and $\alpha$ was confirmed in the E-Log critical regime~\cite{Hu2021}. Shortly afterwards, consistent estimates of $\hat{q}$ and $\alpha$ were obtained for $n=2$ from different realizations of O(2) criticality~\cite{ToldinMetlitski2021extraordinary,Zhang2022Surface,Zou2022Surface,SunQuantum2022,SunClassical2022,Zhang2023} (Table~\ref{tb1}).

\begin{figure}
\includegraphics[height=5.2cm,width=8.8cm]{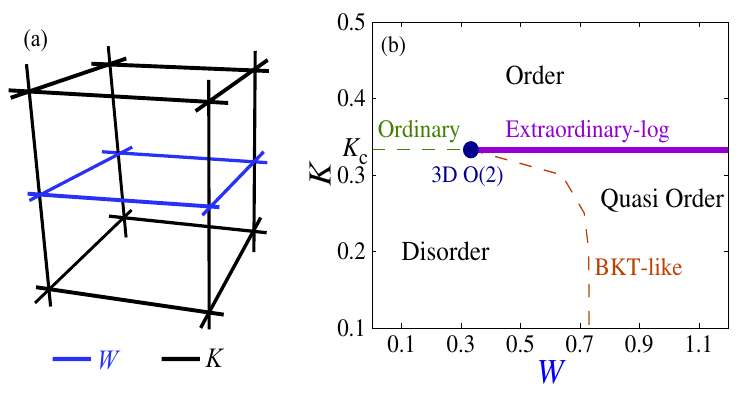}
\caption{A 3D lattice with a plane defect (a) and the phase diagram for the plane defect in 3D Villain model (b). $W$ and $K$ represent the interactions inside and outside the plane defect, respectively. There are quasi-long-range ordered, ordered and disordered phases. The plane-defect critical phenomena include the ordinary, 3D O(2) and E-Log criticality at the bulk critical point $K_c$ as well as the BKT-like transition for $K<K_c$.}~\label{Fig1}
\end{figure}

 \begin{figure*}
\includegraphics[height=5.8cm,width=18cm]{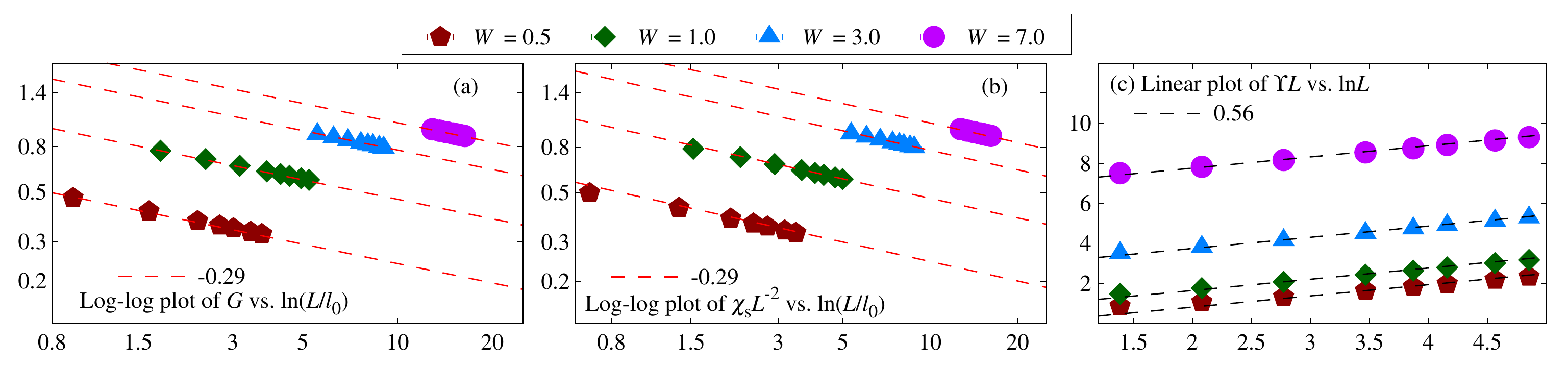}
\caption{The two-point correlation $G$ (a), the scaled susceptibility $\chi_s L^{-2}$ (b) and the scaled helicity modulus $\Upsilon L$ (c) for the E-Log critical phase of the 3D plane-defect Villain model. In panels (a) and (b), the horizontal coordinates are set as ${\rm ln}(L/l_0)$, where the values of $l_0$ are from least-squares fits, and the plots are further made on log-log scales. The slopes $-0.29$ and $0.56$ of dashed lines stand for $-\hat{q}$ and $\alpha$, respectively.}\label{Fig2}
\end{figure*}

\begin{table}
 \begin{center}
 \caption{E-Log critical phases of O($n$) systems. ``BU'' denotes bulk universality, and $\vartriangle$ ($\blacktriangle$) represents the conformity (inconformity) with scaling relation~(\ref{SR_qalpha}).}\label{tb1}
 \begin{tabular}{p{1.4cm}p{2.2cm}p{1.2cm}p{1.4cm}p{0.8cm}p{0.9cm}}
   \hline
   \multicolumn{6}{c}{\textbf{Surface critical phenomena}} \\
   BU& model &  $\hat{q}$ &  $\alpha$ &  $\vartriangle$/$\blacktriangle$ &year \\
   \hline
  \multirow{2}{*}{3D O(3)} &O(3) $\phi^4$~\cite{toldin2020boundary} &  2.1(2)  & 0.15(2) & $\vartriangle$ & 2020  \\
  &O(3) $\phi^4$~\cite{ToldinMetlitski2021extraordinary} &  & 0.190(4) & &2021 \\
   \hline
   \multirow{6}{*}{3D O(2)}&$XY$~\cite{Hu2021} &  0.59(2)  & 0.27(2) & $\vartriangle$ & 2021  \\
  &O(2) $\phi^4$~\cite{ToldinMetlitski2021extraordinary} &  & 0.300(5) & &2021 \\
   &Potts~\cite{Zhang2022Surface}     & 0.60(2) &   & & 2022  \\
  &clock~\cite{Zou2022Surface}  & 0.59(1) &  0.26(2) & $\vartriangle$& 2022  \\
 &Villain~\cite{SunClassical2022}  &  0.58(2) &  0.28(1) & $\vartriangle$ & 2022 \\
  &Potts~\cite{Zhang2023}  &  0.59(3) &   &  &2023 \\
  \hline
   \multicolumn{6}{c}{\textbf{Plane-defect critical phenomena}} \\
    BU& model &  $\hat{q}$ &  $\alpha$ & $\vartriangle$/$\blacktriangle$ &  year \\
        \hline
   \multirow{2}{*}{3D O(2)}  & field theory~\cite{Krishnan2023}   &   & 0.600(10) & & 2023\\
     & Villain, $XY$  &  0.29(2) & 0.56(3) &  $\vartriangle$ & present \\
        \hline
     \multirow{2}{*}{3D O(3)}  & field theory~\cite{Krishnan2023}  &  & 0.540(8) && 2023 \\
    & Heisenberg  &  0.63(3) & 0.33(2) & $\blacktriangle$  & present \\
       \hline
     \multirow{1}{*}{4D O(2)}  &  $XY$  &  0.09(2) & 0.97(7) &  &present \\
    \hline
 \end{tabular}
 \end{center}
 \end{table}

Since the E-Log criticality has been found merely for the SCB of 3D systems, a generalization is essential. Using large-scale Monte Carlo simulations, we provide smoking-gun evidence of the E-Log criticality in the plane defects sitting inside critical 3D Villain, $XY$, Heisenberg and O(6) vector models. The emergence of the E-Log criticality is consistent with a very recent field-theoretic result~\cite{Krishnan2023}. Furthermore, we find that, while the values of $\hat{q}$ and $\alpha$ for $n=2$ are compatible with scaling relation (\ref{SR_qalpha}), violations of this well-established scaling relation are found for $n\geq3$. Another important generalization is to the 4D $XY$ model. Despite the trivial mean-field critical behavior in the bulk, we find that the plane defect can enter the E-Log critical phase via an exotic transition. Note that the O($n$) spin model is perhaps the most important class of models in statistical mechanics and has broad application in condensed-matter physics. In particular, we expect that the E-Log universality in the plane defects would find numerous realizations in the line defects of 2D and 3D quantum models for superfluidity, superconductivity, magnetism, etc.

\noindent{\textcolor{blue}{\it E-Log universality in 3D plane-defect Villain model.}---}
We consider a plane-defect Villain model on the simple-cubic lattice with Hamiltonian $\scrh=\frac{1}{2}\sum_{\bf \langle rr' \rangle} \frac{\scrj^2_{{\bf rr'}}}{C_{{\bf rr'}}}$, where $\scrj_{{\bf rr'}} \in \{0, \pm 1, \pm 2, ...\}$ represents directed flows along the bonds between the nearest neighbors ${\bf r}$ and ${\bf r'}$, and $C_{\bf rr'}$ controls their relative weights. As in the standard Villain model~\cite{chauniversal,WallinSuper,aletcluster,vsmakovuniversal,witczakdynamics,chenuniversal}, the flows are non-divergent and constitute closed directed loops. Periodic boundary conditions are imposed in each of the [100] ($x$), [010] ($y$), and [001] ($z$) directions. To involve a plane defect, we specify a plane that is perpendicular to $z$ direction [Fig.~\ref{Fig1}(a)]. If ${\bf r}$ and ${\bf r'}$ are both in the plane defect, we set $C_{\bf rr'}=W$; otherwise, we let $C_{\bf rr'}=K$. The bulk critical point $K_c$ was located at $K=0.33306704(7)$ with $W=K$~\cite{Xu2019} and falls into 3D O(2) universality class.

We formulate a variant of Prokof'ev-Svistunov worm Monte Carlo algorithm that has two update schemes respectively for entire lattice and plane defect, and is very convenient for the measurements of correlation function and susceptibility in the plane defect; see the supplemental material (SM, which includes~\cite{prokof2001worm} and \cite{PhysRevB.108.L020404,Xu2019,PhysRevB.102.024406}).

In Fig.~\ref{Fig1}(b), we map out a phase diagram. When the bulk is critical ($K=K_c$), a plane-defect transition occurs at $W=K_c$ as $W$ is varied and has the effective thermodynamic renormalization exponent $y_t=1/\nu_{3xy}-1 \approx 0.4885$, with $\nu_{3xy}=0.67183(18)$~\cite{Xu2019} the correlation length exponent of 3D O(2) bulk criticality. This transition differs from the special transition for open surfaces in O(2) systems, which has $y_t=0.58(1)$~\cite{deng2005surface,SunClassical2022}. Additionally, we confirm the field-theoretic prediction~\cite{Diehl1986} of ordinary critical phase for $W<K_c$ at $K=K_c$ and the BKT-like transition with $K<K_c$. The BKT-like transition arises from the essential 2D critical behavior of plane defect in a disordered bulk~\cite{Diehl1986}. For these plane-defect critical phenomena, Monte Carlo results are presented in SM.

\begin{figure}
\centering
\includegraphics[height=8cm,width=8.5cm]{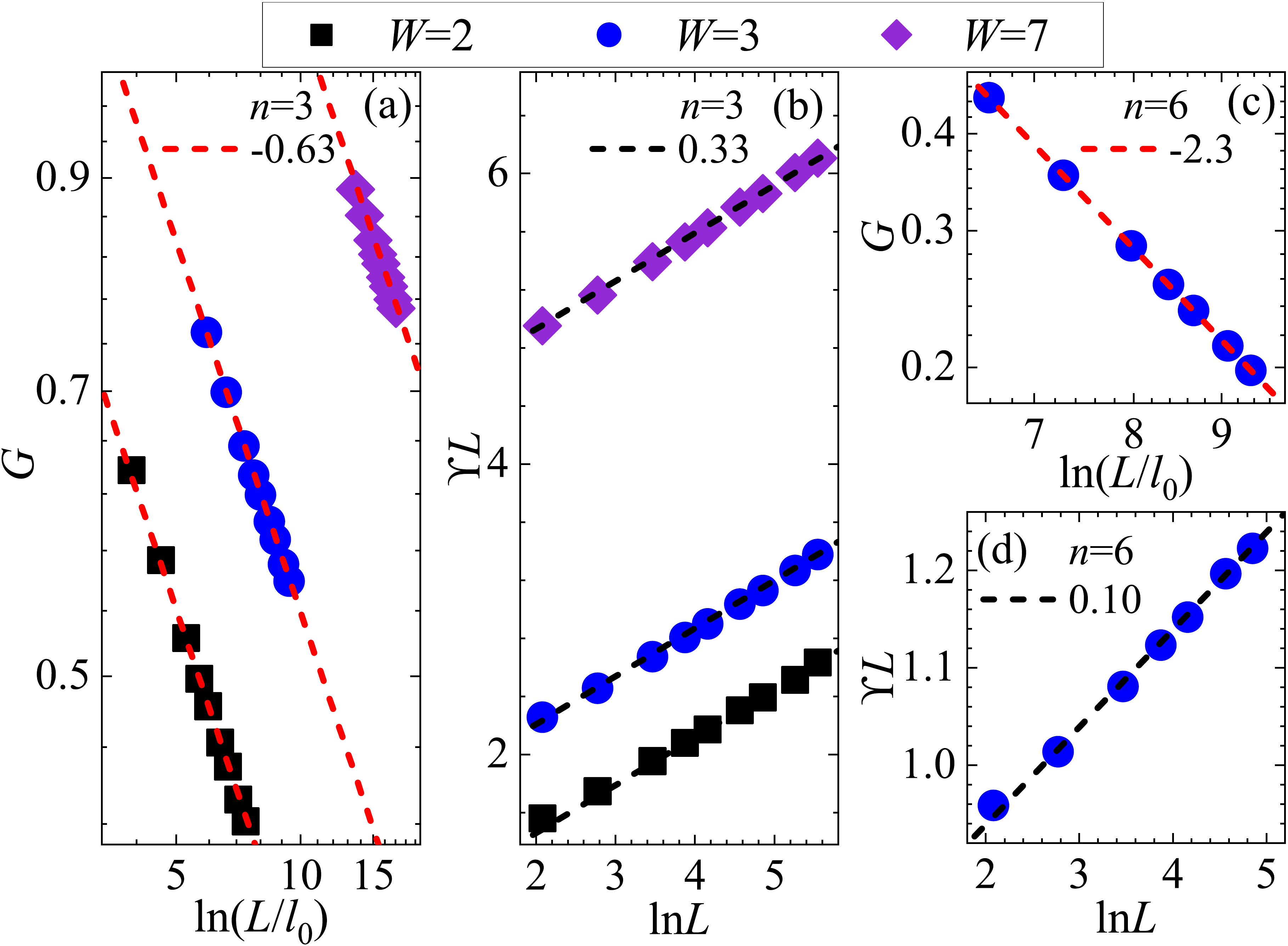}	
\caption{Same as Figs.~\ref{Fig2} (a) and (c) but for the E-Log critical phases of the 3D plane-defect Heisenberg and O(6) vector models.}~\label{Fig3}
\end{figure}

We then explore the strong-$W$ regime with $K=K_c$. Extensive worm-algorithm simulations are performed at $W=0.5$, $1$, $3$ and $7$, with $L$ ranging from $4$ to $128$. At $L=128$, the total number of generated closed loops for a specified $W$ reaches $2.9 \times 10^{10}$.
If the E-Log criticality happens, a fitting ansatz of the large-distance correlation $G$ in the plane defect is $G = a_0[{\rm ln}(L/l_0)]^{-\hat{q}}$, with $a_0$ a non-universal constant and $l_0$ a reference length. Generally speaking, numerical analyses of finite-size scaling (FSS) involving ${\rm ln}L$ are challenging. Such a difficulty can be alleviated at the cost of systematic fits. Throughout this paper, we test scaling ans\"{a}tze against data by least-squares fits. We monitor the evolution of $\chi^2$ by changing the minimum size $L_{\rm min}$ involved. In principle, one searches for the smallest $L_{\rm min}$ relating to the $\chi^2$ per degree of freedom (DoF) $\chi^2/{\rm DoF}=\scro(1)$, which does not decrease drastically upon further increasing $L_{\rm min}$. Practically one prefers the fits with $\chi^2/{\rm DoF} \approx 1$. We should not trust any single fit and conclusions will be made by comparing preferred fits. For each $W$, we find that the estimate of $\hat{q}$ extrapolates to $\hat{q} \approx 0.29$. More precisely, for $W=0.5$, $1$, $3$ and $7$, we obtain $\hat{q}=0.308(2)$, $0.301(2)$, $0.289(5)$ and $0.28(1)$ as well as $\chi^2/{\rm DoF} \approx 2.8$, $0.1$, $2.8$ and $1.5$ respectively, with $L_{\rm min}=16$. Based on these observations, by fixing $\hat{q}=0.29$, we obtain $l_0=3.15(2)$, $0.684(2)$, $0.0153(2)$ and $0.0000100(3)$ as well as $\chi^2/{\rm DoF} \approx 2.4$, $1.4$, $0.8$ and $0.7$, with $L_{\rm min}=48$, $32$, $64$ and $48$, for $W=0.5$, $1$, $3$ and $7$, respectively. Details of fits are given in SM. The consistent estimates of $\hat{q}$ from various $W$ indicate the uniqueness of logarithmic scaling and critical exponent.

Borrowing the insights into FSS from the E-Log criticality of SCB, we obtain the FSS $\chi_s =a_1 L^2 [{\rm ln}(L/l_0)]^{-\hat{q}}$ of the plane-defect susceptibility $\chi_s$, with $a_1$ a non-universal constant. For $W=0.5$, the fit with $L_{\rm min}=16$ yields $\hat{q}=0.322(1)$ and has a huge $\chi^2/{\rm DoF}$ ($\chi^2/{\rm DoF} \approx 7.0$), which reduces to $\chi^2/{\rm DoF} \approx 2.7$ at $L_{\rm min}=32$ with $\hat{q}=0.309(3)$. For $W=1$, $3$ and $7$, we obtain $\hat{q}=0.310(1)$, $0.295(3)$ and $0.29(1)$ as well as $\chi^2/{\rm DoF} \approx 1.7$, $3.2$ and $3.0$, respectively. When $\hat{q}=0.29$ is fixed, we obtain $l_0=4.08(3)$ ($W=0.5$), $0.866(3)$ ($W=1$), $0.01909(9)$ ($W=3$) and $0.0000119(3)$ ($W=7$) with $0.3\lessapprox\chi^2/{\rm DoF}\lessapprox 1.4$. Hence, the estimates of $\hat{q}$ are close to those from $G$.

From the FSS analyses of $G$ and $\chi_s$, we estimate $\hat{q}=0.29(2)$. In Figs.~\ref{Fig2}(a) and (b), by plotting $G$ and $\chi_s L^{-2}$ versus ${\rm ln}(L/l_0)$, where the values of $l_0$ are from fits, the mutually consistent scaling formulae and critical exponent are illustrated.

\begin{figure*}
\includegraphics[height=5.8cm,width=18.3cm]{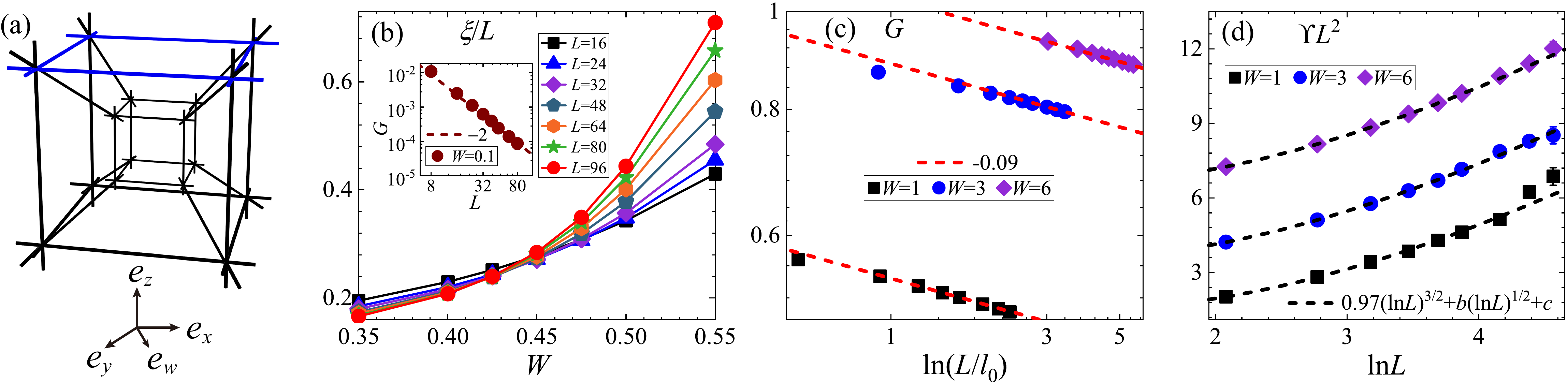}
\caption{Plane-defect criticality of 4D $XY$ model at $K=K_c$. (a) A 4D hypercubic lattice with a plane defect. (b) The scaled second-moment correlation length $\xi/L$ versus $W$. Inset: log-log plot of $G$ versus $L$ at $W=0.1$. (c) Log-log plot of $G$ versus ${\rm ln}(L/l_0)$ for $W=1$, $3$ and $6$, where the values of $l_0$ are from least-squares fits. The slope $-0.09$ of dashed lines stands for $-\hat{q}$. (d) $\Upsilon L^2$ versus ${\rm ln}L$. The dashed lines stand for $\Upsilon L^2=0.97({\rm ln}L)^{3/2}+b ({\rm ln}L)^{1/2}+c$, where $b$ and $c$ are from least-squares fits.}\label{Fig4}
\end{figure*}

We analyze the helicity modulus $\Upsilon$, which is defined through the fluctuations of winding number $\mathcal{W}_x$ of directed flows along a periodic direction (say $x$ direction), namely $\Upsilon=\langle \mathcal{W}_x^2 \rangle/L$. For the E-Log criticality in a 3D system, $\Upsilon$ scales as $\Upsilon L \sim {\rm ln}L$. This behavior is roughly illustrated by the data in Fig.~\ref{Fig2}(c). We perform least-squares fits to $\Upsilon L = \alpha ({\rm ln}L)+b+cL^{-\omega}$, where $\alpha$ can be universal, $\omega$ denotes the exponent of leading finite-size corrections, whereas $b$ and $c$ are non-universal. Unlike the scaling form $\Upsilon L = 2\alpha ({\rm ln}L)+b+cL^{-\omega}$ that applies to SCB involving two open surfaces, the prefactor is $\alpha$ due to the uniqueness of plane defect. We look into the fits using $\omega=0.789$~\cite{Criticalexponents1998Guida,Monte2019Hasenbusch} or $\omega=1$, and find that the inclusion of correction term stabilizes fits. For $W=0.5$, $1$, $3$ and $7$, we obtain $\alpha=0.555(3)$, $0.562(4)$, $0.580(6)$ and $0.57(1)$ as well as $\chi^2/{\rm DoF} \approx 0.2$, $0.7$, $2.0$ and $2.2$ respectively, with $L_{\rm min}=16$. Comparing all preferred fits (SM), we estimate $\alpha=0.56(3)$.

The universal results of $\hat{q}$ and $\alpha$ from different $W$ prove the universality of the E-Log criticality.
Furthermore, the values of $\hat{q}$ and $\alpha$ cannot be related to any known E-Log criticality, indicating a new universality class. In SM, it is demonstrated that $\hat{q}$ and $\alpha$ from direct fits are consistent with those from the conversions according to Eq.~(\ref{SR_qalpha}), and such consistency is even more obvious at $\chi^2/{\rm DoF} \approx 1$. Thus, scaling relation~(\ref{SR_qalpha}) is validated for the present E-Log universality.

\noindent{\textcolor{blue}{\it E-Log criticality in 3D plane-defect O($n$) vector models.}---} We consider the plane-defect O($n$) vector models with Hamiltonian $\scrh = - {\sum_{\langle {\bf r}{\bf r'} \rangle}} C_{{\bf r}{\bf r'}} \vec{S}_{\bf r} \cdot \vec{S}_{\bf r'}$, where $\vec{S}_{\bf r}$ represents $n$-component unit-vector spins. Using Wolff cluster Monte Carlo algorithm~\cite{WolffCollective1989}, we sample the helicity modulus $\Upsilon=\frac{1}{L^d}(\frac{2}{n}\langle E \rangle-\frac{2}{n(n-1)}\sum_{a<b}\langle (T^{(a,b)}_{\bf{e}_x})^2\rangle)$ with $E = \sum_{\bf{r}} C_{{\bf r}({\bf{r}+\bf{e}_x})}  \vec{S}_{\bf{r}}\cdot\vec{S}_{\bf{r}+\bf{e}_x}$ and	
$T^{(a,b)}_{\bf{e}_x}= \sum_{\bf{r}} C_{{\bf r}({\bf{r}+\bf{e}_x})} (S_{\bf{r}}^a S_{\bf{r}+\bf{e}_x}^b - S_{\bf{r}}^b S_{\bf{r}+\bf{e}_x}^a)$, where ($S_{\bf{r}}^a$, $S_{\bf{r}}^b$) represents pairs of components of $\vec{S}_{\bf r}$, as well as the two-point correlation $G=\langle \vec{S}_{\bf{r}} \cdot \vec{S}_{\bf{r'}} \rangle$ with ${\bf r'}-{\bf r}=(L/2, 0)$ in the plane defect.

We consider the $n=2$ ($XY$) case on the simple-cubic lattice with $K=K_c=1/2.2018441$, where $K_c$ is the bulk critical point~\cite{Xu2019}, and simulate at $W=1$, $3$ and $7$. For $G$ and $\Upsilon$, we perform FSS analyses using the scaling ans\"{a}tze for the E-Log criticality. In SM, using a specially designed $\chi^2$ test, we confirm that $\hat{q}$ and $\alpha$ agree with those from the plane-defect Villain model. Thus far, all uncovered E-Log critical phenomena conform to scaling relation (\ref{SR_qalpha}).

We now study the E-Log universality for the $n=3$ (Heisenberg) case on the simple-cubic lattice. We start by simulating the $W=K$ case up to $L=384$ and obtain $K_c \approx 0.69300288$, which surpasses the most accurate result $K_c=0.693003(2)$ in literature~\cite{deng2005surface}. We then simulate at $W=2$, $3$ and $7$ with $K=0.69300288$, and confirm the existence of E-Log criticality (SM). Furthermore, we find the universality of $\hat{q}$ and $\alpha$ with $\hat{q}=0.63(3)$ and $\alpha=0.33(2)$, which violate scaling relation (\ref{SR_qalpha}). With the estimated $\hat{q}$ and $\alpha$, the FSS of $G$ and $\Upsilon$ are displayed in Figs.~\ref{Fig3}(a) and (b), respectively.

Due to the upper bound $n_c \approx 5$, the E-Log criticality does not exist for $n>5$ in the open surfaces of 3D systems~\cite{padayasi2022extraordinary}. By contrast, for the plane-defect O(6) vector model, we find the E-Log criticality with $\hat{q}=2.3(1)$ and $\alpha=0.10(1)$ at $K=K_c=1.428653$~\cite{Liu2023} and $W=3$ [Figs.~\ref{Fig3}(c) and (d)], which again violates scaling relation (\ref{SR_qalpha}).

\noindent{\textcolor{blue}{\it Exotic plane-defect transition and E-Log universality in 4D $XY$ model.}---} The E-Log criticality has merely been found in the plane defects and open surfaces of 3D systems. Is there an E-Log universality class for any other spatial dimension? We consider the plane-defect $XY$ model on 4D hypercubic lattices [Fig.~\ref{Fig4}(a)] with $K=K_c$, for which one has $K_c=1/3.314437$~\cite{lv2019}.

At $W=K_c$, the effective plane-defect thermodynamic renormalization exponent $y_t$ vanishes, since $y_t=1/\nu_{4xy}-2$ and $\nu_{4xy}=1/2$ apply, and the interaction enhancement can be exactly marginal, marginally relevant or marginally irrelevant. The scaled second-moment correlation length $\xi/L$ indicates a transition at $W_c \approx 0.41$ by the deviation from scale invariance for $W>W_c$ [Fig.~\ref{Fig4}(b)]. In SM, we perform systematic FSS analyses using various scaling ans\"{a}tze: a standard scaling with $L^{y_t}$ (with or without logarithmic corrections) and a scaling with $y_t=0$ but involving ${\rm ln}L$. The estimates of $W_c$ from preferred least-squares fits are compatible with $W_c=0.41(2)$. Hence, our results reveal a transition at $W_c$, which is compatible with the marginally irrelevant scenario, since $W_c$ is significantly larger than $K_c$. Similar to the FSS at $W=K_c$, $G$ scales as $G \sim L^{-2}$ at $W=0.1$, indicating a critical behavior for $W<W_c$ governed by the Gaussian fixed point.

For $W>W_c$, we find the E-Log criticality. Figure~\ref{Fig4}(c) shows, for $W=1$, $3$ and $6$, that $G$ scales as $G \sim [{\rm ln}(L/l_0)]^{-\hat{q}}$ with the universal exponent $\hat{q}=0.09(2)$. Divergence of $\Upsilon L^2$ upon increasing $L$ is inferred from Fig.~\ref{Fig4}(d). Using least-squares fits, we find that $\Upsilon$ scales as $\Upsilon L^2 [{\rm ln}(L/l_0)]^{-1/2} \sim \alpha {\rm ln}(L/l_0)$ with the universal parameter $\alpha=0.97(7)$. In this FSS formula, the left-hand side relates to the FSS $\Upsilon_{\rm 4D} \sim L^{-4} (\xi_{\rm 4D})^2 \sim L^{-2}[{\rm ln}(L/l_0)]^{1/2}$ of 4D bulk criticality with the exponents $-2$ and $1/2$ for leading scaling and logarithmic correction respectively~\cite{lv2019}, whereas the E-Log universality accounts for $\alpha {\rm ln}(L/l_0)$ in the right-hand side.

\noindent{\textcolor{blue}{\it Summary and discussions.}---} We study the plane-defect criticality of the O($n$) model with $n=2$, $3$ and $6$ in three dimensions and $n=2$ in four dimensions, and obtain convincing evidence of the E-Log criticality for each situation. In three dimensions, the E-Log criticality for $n\geq3$ violates scaling relation~(\ref{SR_qalpha}), which holds for $n=2$. For a plane defect in the critical 4D $XY$ system, the presence of E-Log universality and exotic transition is also evidenced. These findings significantly expand the current understanding of E-Log criticality.

The study for 3D and 4D plane-defect systems is a remarkable step toward exploring E-Log criticality in generic O($n$) systems with effective interactions. The hints for such a generalization also come from the logarithmic forms of correlators in certain 2D O($n$) loop models~\cite{nahum2013loop,wang2015completely}.

Our findings can be realized with {\it emergent} O($n$) symmetry or O($n$)-symmetric Hamiltonian with $n \geq 2$. Due to classical-quantum correspondence, our results further indicate the E-Log universality for the line defects in 2D and 3D quantum O($n$) systems~\cite{Sachdev2011QPT,02GreinerQuantum,16BaierExtended,20YangCooling,zhang2023superconducting,14MerchantQuantum}. Besides, the conformal field theory of plane-defect criticality is currently a subject of intensive research~\cite{Trepanier2023,RavivMoshea2023,Giombi2023,bolla2023defects}.

\begin{acknowledgments}
{\it{Acknowledgment.-}} This work was supported by the National Natural Science Foundation of China (Grant Nos. 12275002, 12275263, 11975024), Innovation Program for Quantum Science and Technology (Grant No. 2021ZD0301900), Natural Science Foundation of Fujian province of China
(under Grant No. 2023J02032), and National Key R\&D Program of China (Grant No. 2018YFA0306501).
\end{acknowledgments}

\bibliography{papers}

\end{document}


\title{Supplemental Material for \\  ``Extraordinary-log Universality of Critical Phenomena in Plane Defects''}
\author{Yanan Sun}
\thanks{Y.S. and M.H. contributed equally to this work.}
\author{Minghui Hu}
\thanks{Y.S. and M.H. contributed equally to this work.}
\affiliation{Department of Physics, Anhui Key Laboratory of Optoelectric Materials Science and Technology, Key Laboratory of Functional Molecular Solids, Ministry of Education, Anhui Normal University, Wuhu, Anhui 241000, China}
\author{Youjin Deng}
\email{yjdeng@ustc.edu.cn}
\affiliation{National Laboratory for Physical Sciences at Microscale, University of Science and Technology of China, Hefei, Anhui 230026, China}
\affiliation{Department of Modern Physics, University of Science and Technology of China, Hefei, Anhui 230026, China}
\author{Jian-Ping Lv}
\email{jplv2014@ahnu.edu.cn}
\affiliation{Department of Physics, Anhui Key Laboratory of Optoelectric Materials Science and Technology, Key Laboratory of Functional Molecular Solids, Ministry of Education, Anhui Normal University, Wuhu, Anhui 241000, China}
\begin{abstract}
In this Supplemental Material, we introduce the update schemes of the worm Monte Carlo algorithm for the three-dimensional (3D) plane-defect Villain model. We analyze the Monte Carlo data for the extraordinary-log (E-Log) critical phases of the 3D plane-defect Villain and $XY$ models. We illustrate the ordinary critical phase, multi-critical point and Berezinskii-Kosterlitz-Thouless-like transition of the 3D plane-defect O(2) model. Subsequently, we precisely locate the bulk critical point for the 3D Heisenberg model and reveal the E-Log critical phases of the 3D plane-defect Heisenberg and O(6) vector models. Finally, we study the critical 4D plane-defect $XY$ model, which features E-Log universality associated with an exotic transition.
\end{abstract}

\maketitle

\tableofcontents

\section{Update schemes of worm Monte Carlo algorithm for the 3D plane-defect Villain model}
To simulate the three-dimensional (3D) plane-defect Villain model, we formulate an unbiased worm algorithm inspired by Ref.~\cite{PS2001} in an extended state space of directed flows, where each state has a pair of source-sink points [referred to as Ira ($I$) and Masha ($M$)]. The worm algorithm features a biased random walk of $I$ and $M$, which obeys the detailed balance. A core part of the algorithm relies on iteratively carrying out two update schemes, \textbf{Algorithm} \textbf{1} and \textbf{Algorithm} \textbf{2}, which run in the entire simple-cubic lattice and the plane defect, respectively.

Once $I$ and $M$ collide ($I=M$), a closed loop of directed flows is created. After producing a permanent quantity of closed loop(s), the update process shifts between \textbf{Algorithm} \textbf{1} and \textbf{Algorithm} \textbf{2}. Even though \textbf{Algorithm} \textbf{1} guarantees ergodicity, we include \textbf{Algorithm} \textbf{2} to perform efficient simulations in the plane defect, which are useful for sampling quantities that characterize plane-defect criticality.

Using \textbf{Algorithm} \textbf{2}, we extract the two-point correlation $g(\delta x, \delta y)$ [$g(0,0)\equiv 1$] in plane defect  from the distribution of the distances $(\delta x, \delta y)$ between $I$ and $M$. We define the large-distance two-point correlations by $G=[g(0,L/2)+g(L/2,0)]/2$ and $G'=[g(0,L/4)+g(L/4,0)]/2$. Meanwhile, the susceptibility $\chi_s$ of plane defect is sampled via $\chi_s=\langle n_s \rangle$, where $n_s$ is the number of the movements of $I$ and $M$ between consecutive hits to the original state space of plane-defect Villain model.

\begin{algorithm}[H]
\caption{Update in simple-cubic lattice}
\label{alg:p1}
\begin{algorithmic}
\item 1. If $I=M$, randomly pick up a lattice site $I'$ in the simple-cubic lattice, with the probability $1/L^3$ for each site. Let $I=M=I'$, ${\rm sign}(I)=1$, ${\rm sign}(M)=-1$.
\item 2. Interchange $I \leftrightarrow M$ and ${\rm sign}(I) \leftrightarrow {\rm sign}(M)$ with the probability $1/2$.
\item 3. Randomly pick up a neighbor $I_\scrn$ of $I$, with the probability $1/6$ for each of the neighbors.
\item 4. Propose to simultaneously move $I \rightarrow I_\scrn$ and update the flow $\scrj_{II_\scrn}$ to $\scrj'_{II_\scrn}$:
\begin{equation}
\scrj'_{II_\scrn}=\scrj_{II_\scrn}+{\rm d_f}(I \rightarrow I_\scrn){\rm sign}(I), \nonumber
\end{equation}
where ${\rm d_f}(I \rightarrow I_\scrn)=\pm 1$ is a fixed parameter, quantifying the direction of flow along bond-$II_\scrn$.
\item 5. Accept the proposed change with the probability
\begin{equation}
p=\left\{\begin{array}{cc}
\min[1, e^{\frac{-(\scrj'^2_{II_\scrn}-\scrj^2_{II_\scrn})}{2W}}] &  \,\, I \,\, {\rm and} \, I_\scrn \,\in {\rm plane \,\, defect},  \nonumber \\
\min[1, e^{\frac{-(\scrj'^2_{II_\scrn}-\scrj^2_{II_\scrn})}{2K}}] & \,\,  I \,\, {\rm or} \, I_\scrn \, \notin {\rm plane \,\, defect}. \nonumber
\end{array}\right.
\end{equation}
\end{algorithmic}
\end{algorithm}

\begin{algorithm}[H]
\caption{Update in plane defect}
\label{alg:p2}
\begin{algorithmic}
\item 1. If $I=M$, randomly pick up a lattice site $I'$ in the plane defect, with the probability $1/L^2$ for each site. Let $I=M=I'$, ${\rm sign}(I)=1$, ${\rm sign}(M)=-1$.
\item 2. The same as \textbf{Algorithm 1}.
\item 3. Randomly pick up a neighbor $I_\scrn$ of $I$ in the plane defect, with the probability $1/4$ for each of the neighbors.
\item 4. The same as \textbf{Algorithm 1}.
\item 5. Accept the proposed change with the probability
\begin{equation}
\label{eq6}
p=\min[1, e^{\frac{-(\scrj'^2_{II_\scrn}-\scrj^2_{II_\scrn})}{2W}}]. \nonumber
\end{equation}
\end{algorithmic}
\end{algorithm}

\section{E-Log critical phase of the 3D plane-defect Villain model}

We simulate the extraordinary-log (E-Log) critical phase of the 3D plane-defect Villain model using the worm algorithm, with the lattice sizes $L=4$, $8$, $16$, $32$, $48$, $64$, $96$ and $128$. We consider the parameters $W=0.5$, $1$, $3$ and $7$ with $K=K_c=0.33306704$. The number of generated closed loops ranges from $9.1 \times 10^8$ to $7.3 \times 10^9$ for $L \leq 32$ and ranges from $1.1 \times 10^{10}$ to $2.9 \times 10^{10}$ for $48 \leq L \leq 128$. The first one sixth of the generated closed loops are utilized for the thermalization in each Markov chain.

We analyze the finite-size scaling (FSS) for the E-Log critical phase. According to a standard criterion, we prefer the fits with $\chi^2/{\rm DoF} \approx 1$ and conclude by comparing the fits that are stable against gradually increasing $L_{\rm min}$, which is the size of the minimum system involved. To supplement the presentation in the main text, we now provide more details about the FSS analyses. We focus on the two-point correlation $G$ and the susceptibility $\chi_s$ for the plane defect as well as the helicity modulus $\Upsilon$. The data of these quantities are fitted to the scaling formulae
 \begin{equation}\label{seG}
 G=a_0 [{\rm ln}(L/l_0)]^{-\hat{q}},
 \end{equation}
  \begin{equation}\label{sechi}
\chi_s=a_1 L^2 [{\rm ln}(L/l_0)]^{-\hat{q}}
 \end{equation}
 and
\begin{equation}\label{seUpsilon}
\Upsilon L =\alpha ({\rm ln}L)+ b+ c L^{-\omega},
\end{equation}
respectively, where $\hat{q}$ and $\alpha$ are somewhat universal, and $\omega$ is a correction exponent. $l_0$, $a_0$, $a_1$, $b$ and $c$ represent non-universal constants. The results of least-squares fits are presented in Tables~\ref{extG}, \ref{extChi} and \ref{extrho}, for $G$, $\chi_s$ and $\Upsilon$, respectively.

\begin{figure*}
\includegraphics[height=10cm,width=10cm]{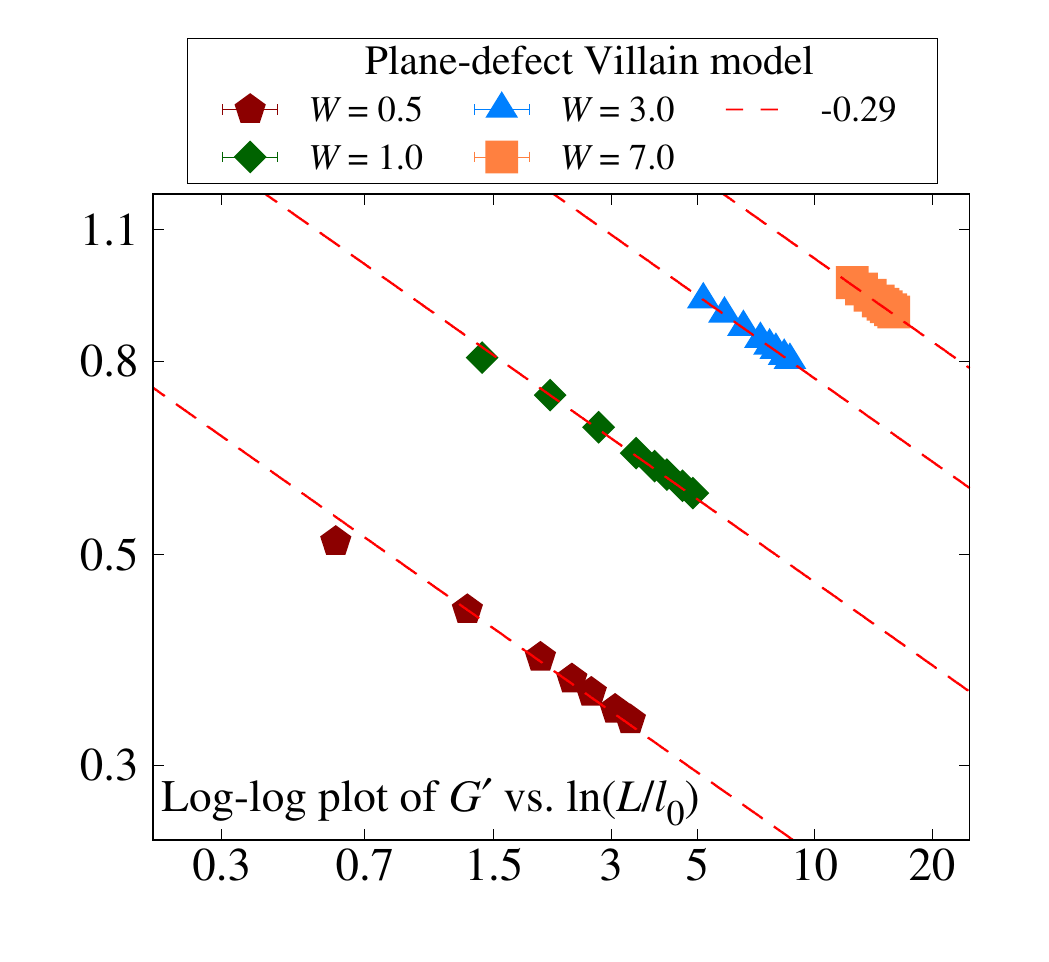}
\caption{Two-point correlation $G'$ versus ${\rm ln}(L/l_0)$ for the E-Log critical phase of the 3D plane-defect Villain model, where $l_0=4.44$ ($W=0.5$), $0.987$ ($W=1$), $0.02261$ ($W=3$) and $0.000015$ ($W=7$) are taken from preferred least-squares fits. The plot is further made on a log-log scale. The slope $-0.29$ of dashed lines stands for $-\hat{q}$.}~\label{FigS1}
\end{figure*}

\begin{figure*}
\includegraphics[height=8cm,width=10cm]{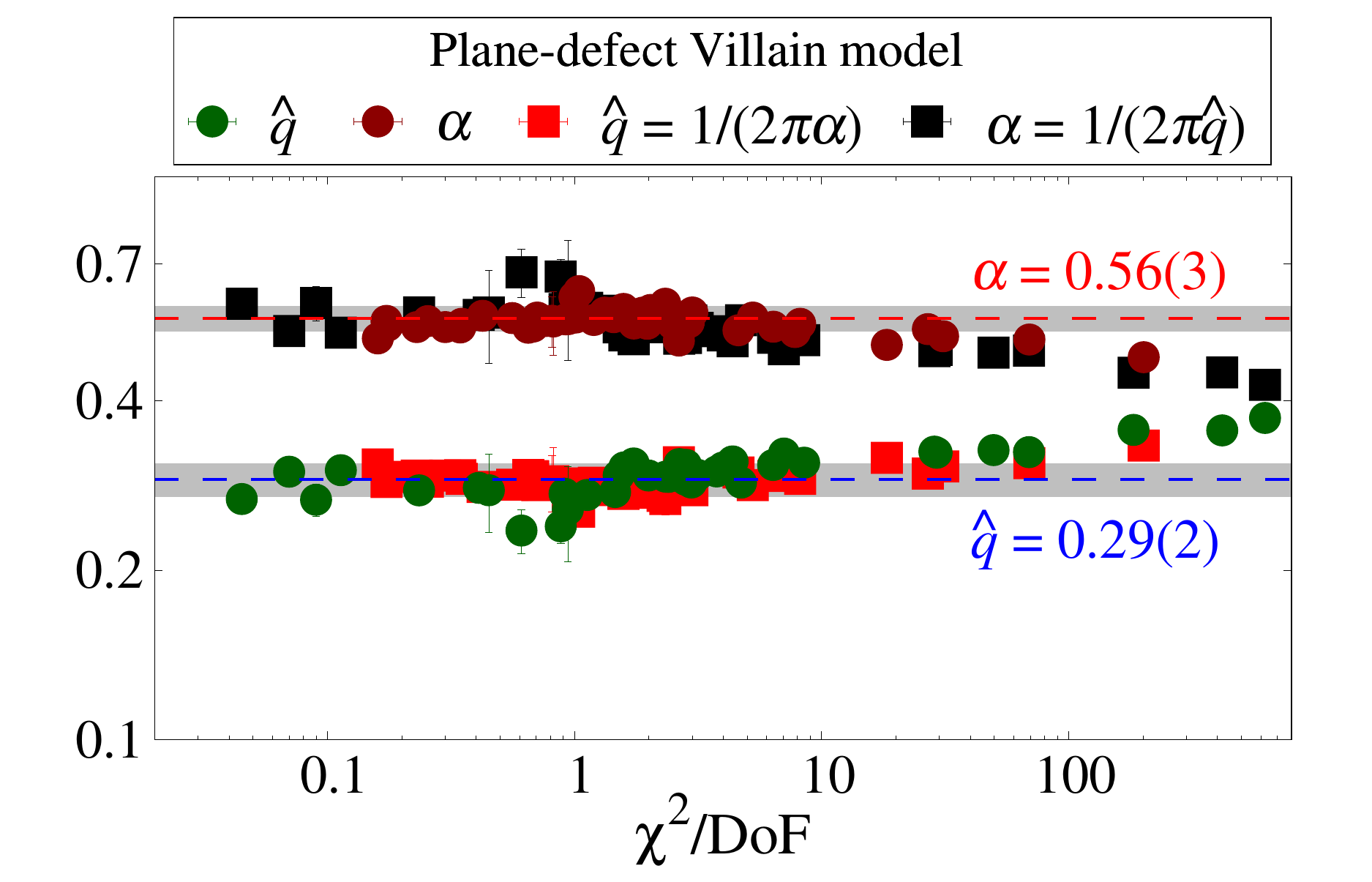}
\caption{Test of the scaling relation $\alpha \hat{q}=1/(2\pi)$ for the E-Log critical phase of the 3D plane-defect Villain model. Circles denote the results from least-squares fits to FSS formulae, whereas squares denote the results converted from fitting results via $\alpha \hat{q}=1/(2\pi)$. The estimated critical exponent $\hat{q}=0.29(2)$ and universal parameter $\alpha=0.56(3)$ are denoted by the shadows centered at the blue and red lines, respectively.}\label{FigS2}
\end{figure*}

In addition to the quantities discussed in the main text, we take into consideration the two-point correlation $G'$. For the E-Log critical phase at $W=0.5$, $1$, $3$ and $7$, we perform FSS analyses according to
\begin{equation}\label{fitseGp}
G'=a_0 [{\rm ln}(L/l_0)]^{-\hat{q}}.	
\end{equation}
The results of fits are summarized in Table~\ref{extGp}. Using the results of $l_0$ from preferred fits, we plot $G'$ versus ${\rm ln}(L/l_0)$ in Fig.~\ref{FigS1}. It is found that the leading FSS behavior of $G'$ is similar to that of $G$.

We analyze the scaling relation $\alpha \hat{q}=1/(2\pi)$ between $\hat{q}$ and $\alpha$. Figure~\ref{FigS2} displays the above-obtained fitting results of $\hat{q}$ and $\alpha$ versus $\chi^2/{\rm DoF}$. Next, from each of the estimates of $\hat{q}$ and $\alpha$, via the equation $\alpha \hat{q}=1/(2\pi)$, we obtain an estimate of $\alpha$ and $\hat{q}$, respectively. The estimates converted through the equation are also included in Fig.~\ref{FigS2}. We note that the results of $\hat{q}$ and $\alpha$ from direct fits and conversions are consistent. For $\chi^2/{\rm DoF} \approx 1$, such consistency is obvious. Hence, we obtain strong evidence of the scaling relation $\alpha \hat{q}=1/(2\pi)$ for the E-Log universality of plane-defect Villain model.

\section{E-Log critical phase of the 3D plane-defect $XY$ model}

We simulate the E-Log critical phase of the 3D plane-defect $XY$ model using Wolff cluster algorithm, with the lattice sizes $L=8$, $16$, $32$, $48$, $64$, $96$ and $128$. We consider the coupling strengths $W=1$, $3$ and $7$ with $K=K_c=1/2.2018441$. The number of Wolff steps ranges from $1.9 \times 10^8$ to $7.7 \times 10^8$ for $L \leq 32$ and ranges from $2.5 \times 10^8$ to $1.2 \times 10^9$ for $48 \leq L \leq 128$. The first one sixth of Wolff Monte Carlo steps are utilized for the thermalization in each Markov chain.

\begin{figure*}
\includegraphics[height=9.5cm,width=10cm]{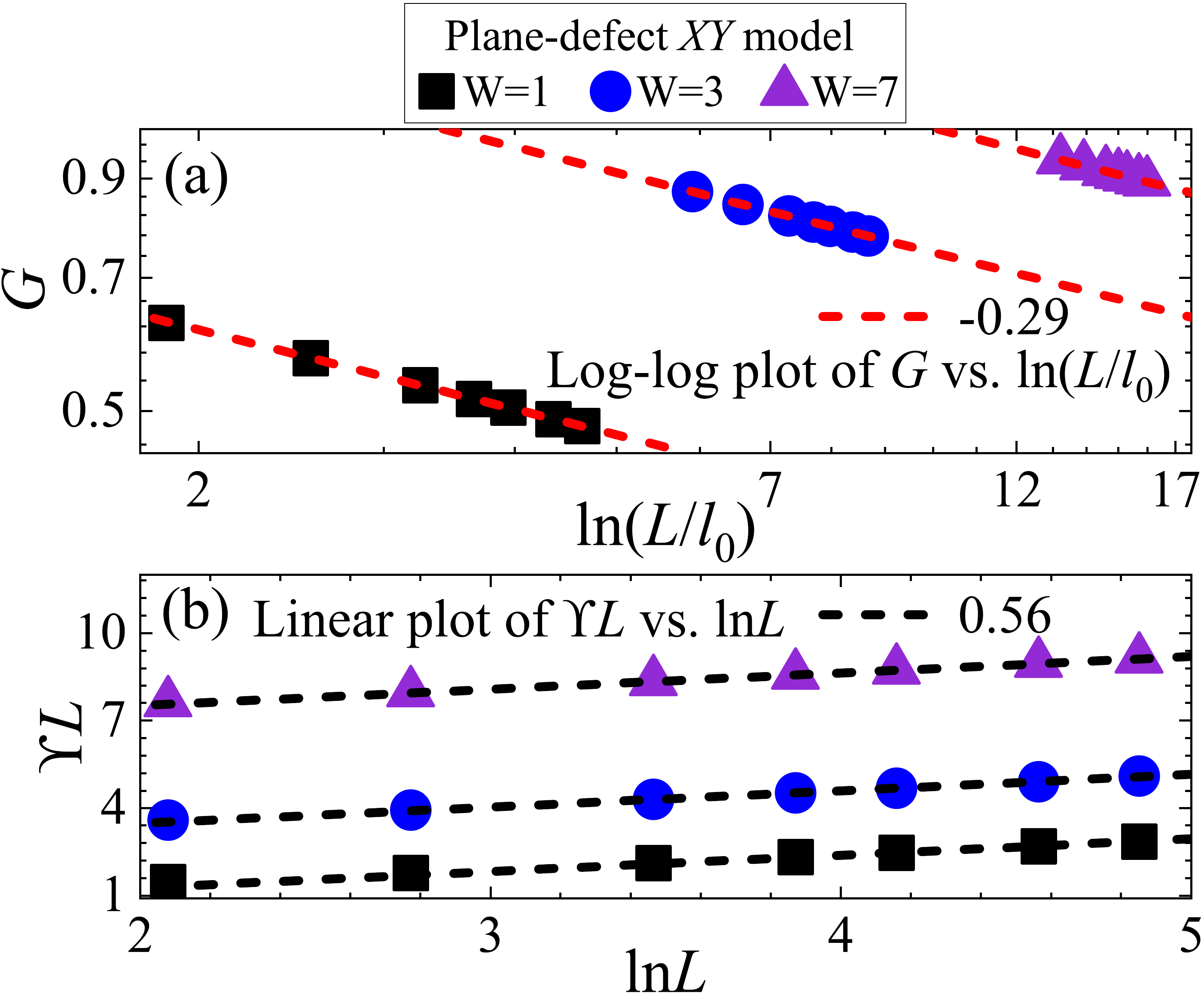}
\caption{The two-point correlation $G$ (a) and the scaled helicity modulus $\Upsilon L$ (b) for the E-Log critical phase of the 3D plane-defect $XY$ model. In panel (a), the horizontal coordinate is written as ${\rm ln}(L/l_0)$, where $l_0=1.240$ ($W=1$), $0.0219$ ($W=3$) and $0.0000146$ ($W=7$) are taken from preferred least-squares fits, and the plot is further made on a log-log scale. The slopes $-0.29$ and $0.56$ of dashed lines stand for $-\hat{q}$ and $\alpha$, respectively.}\label{FigS3}
\end{figure*}

\begin{figure*}
\includegraphics[height=10cm,width=10cm]{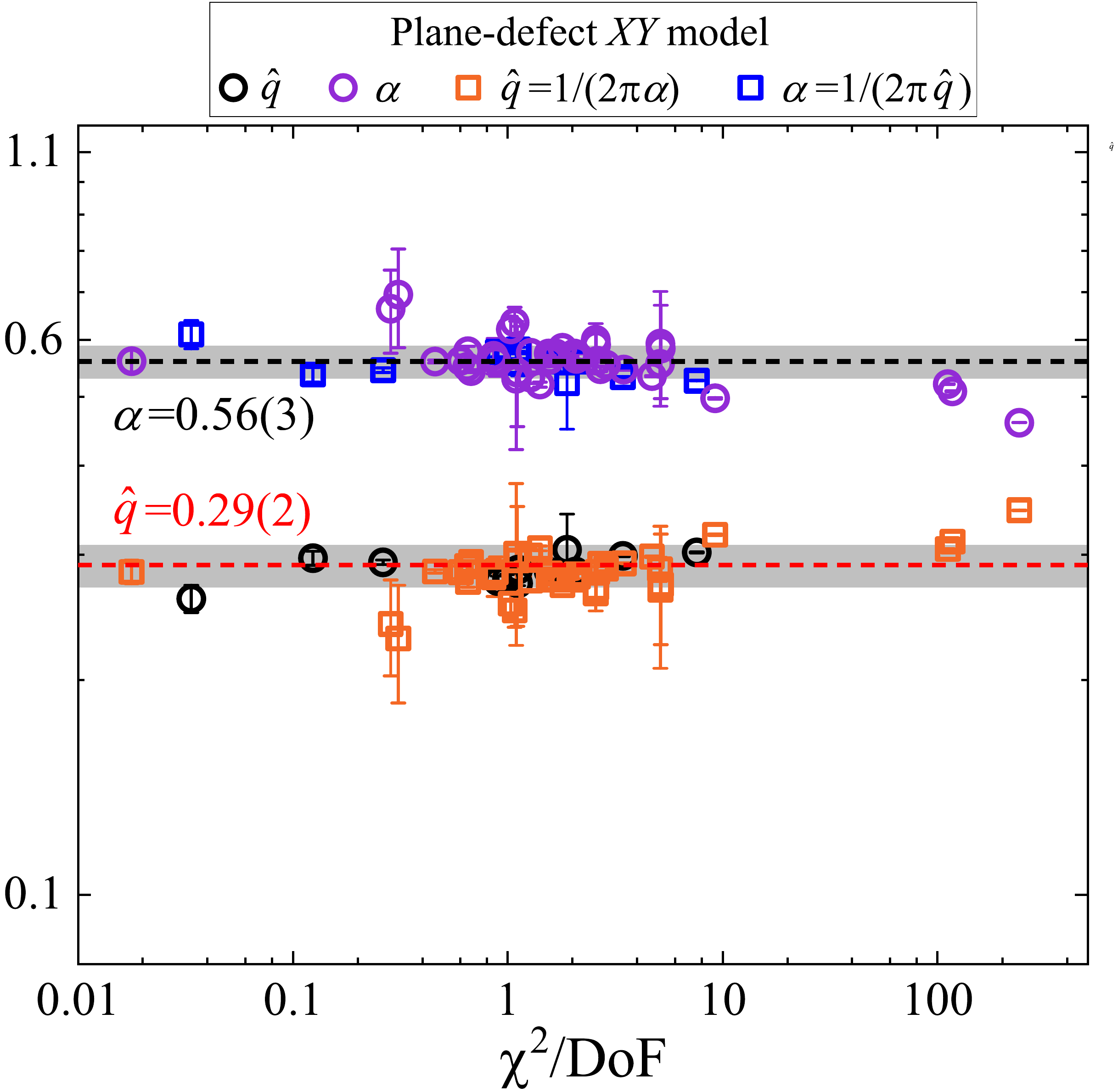}
\caption{Test of the scaling relation $\alpha \hat{q}=1/(2\pi)$ for the E-Log critical phase of the 3D plane-defect $XY$ model. Circles denote the results from least-squares fits to FSS formulae, whereas squares denote the results converted from fitting results via $\alpha \hat{q}=1/(2\pi)$. The estimated critical exponent $\hat{q}=0.29(2)$ and universal parameter $\alpha=0.56(3)$ are denoted by the shadows centered at the red and black lines, respectively.}\label{FigS4}
\end{figure*}

The Monte Carlo data of $G$ and $\Upsilon$ are analyzed according to Eqs.~(\ref{seG}) and (\ref{seUpsilon}), respectively. Tables~\ref{extGXY} and \ref{extUxy} summarize the outcomes of least-squares fits. The FSS of $G$ and $\Upsilon$ are demonstrated by Fig.~\ref{FigS3}. Moreover, using the results of $\hat{q}$ and $\alpha$ from least-squares fits, the scaling relation $\alpha \hat{q}=1/(2\pi)$ is verified by a specially designed $\chi^2$ test, which is displayed by Fig.~\ref{FigS4}.

\section{Ordinary critical phase, multi-critical point and Berezinskii-Kosterlitz-Thouless-like transition of the 3D plane-defect O(2) model}

We study the ordinary critical phases of the 3D plane-defect O(2) models. For the plane-defect Villain model, we find that the FSS of $\chi_s$ at $W=0.3$ is compatible with the scaling form $\chi_s= a L^{2y_h-2}(1+bL^{-\omega})+c$, where $2y_h-2=-0.4572$ is the magnetic renormalization exponent of ordinary universality class~\cite{ParisenToldin2023}, and the correction exponent $\omega=0.789$ of 3D O(2) critical universality is adopted. $a$, $b$ and $c$ are non-universal constants. We obtain $\chi^2/{\rm DoF} \approx 1.9$ and $2.8$ with $L_{\min}=16$ and $32$, respectively. For the plane-defect $XY$ model, we sample $\chi_s$ in spin representation by $\chi_s=\frac{1}{L^2}\langle |\sum_{\bf r}\vec{S}_{\bf r}|^2 \rangle$, where the summation runs over sites in the plane defect. At $W=0.2$, we obtain $2y_h-2=-0.44(1)$ and $-0.45(5)$ with $\chi^2/{\rm DoF} \approx 1.4$ and $2.0$, for $L_{\min}=8$ and $16$, respectively. These results are consistent with the critical exponent of ordinary universality class.

\begin{figure*}
\includegraphics[height=8cm,width=12cm]{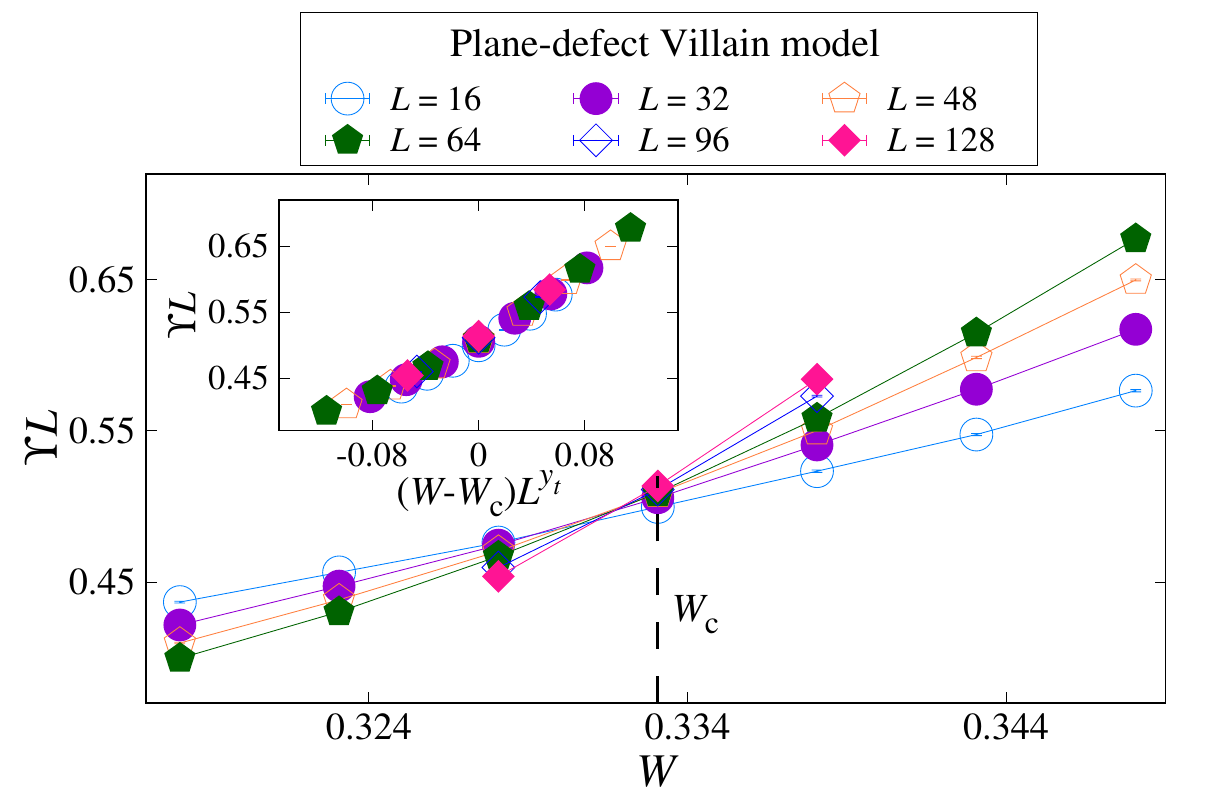}
\caption{The scaled helicity modulus $\Upsilon L$ versus $W$ for the 3D plane-defect Villain model at $K=K_c$. Inset: $\Upsilon L$ versus $(W-W_c)L^{y_t}$ with $W_c=0.33306704$ and $y_t=1/0.67183-1$.}\label{FigS5}
\end{figure*}

We study the multi-critical point of the plane-defect Villain model. This multi-critical point corresponds to a phase transition occurring at $K=W=K_c$ as $W$ is varied. The transition possibly has the effective thermodynamic renormalization exponent $y_t=1/\nu_{3xy}-1$, with $\nu_{3xy}$ the correlation length exponent of 3D O(2) model. Indeed, Fig.~\ref{FigS5} shows the scale-invariant behavior of $\Upsilon L$ at $W=K_c$, which indicates a phase transition. It further shows, around $W=K_c$, that the critical scaling behavior of $\Upsilon L$ is characterized by the exponent $y_t=1/\nu_{3xy}-1$, as $\nu_{3xy}$ takes a literature result $\nu_{3xy}=0.67183(18)$~\cite{Xu2019}. Hence, the transition belongs to 3D O(2) bulk universality class.

\begin{figure*}
\includegraphics[height=10cm,width=12cm]{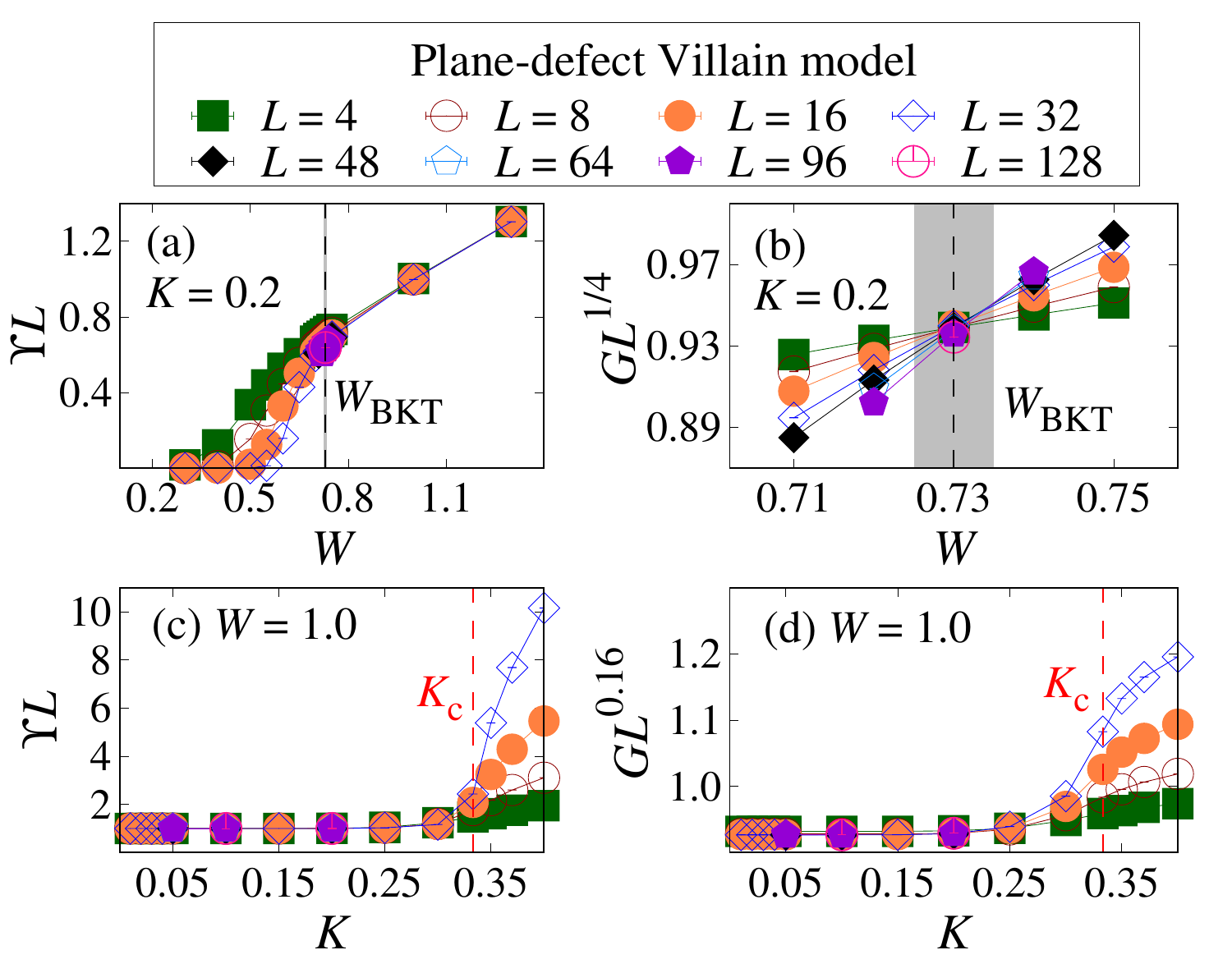}
\caption{Critical phenomena of the 3D plane-defect Villain model with $K \leq K_c$. The scaled helicity modulus $\Upsilon L$ (a) and the scaled two-point correlation $G L^{1/4}$ (b) versus $W$ for $K=0.2$. $\Upsilon L$ (c) and $G L^{0.16}$ (d) versus $K$ for $W=1.0$. Dashed lines represent phase transition points.}\label{FigS6}
\end{figure*}

We explore the critical phenomena of the plane-defect Villain model for $K<K_c$. Figure~\ref{FigS6}(a) shows $\Upsilon L$ versus $W$ at $K=0.2$. For $W \gtrapprox 0.7$, $\Upsilon L$ tends to be independent of $L$ in the $L \rightarrow \infty$ limit and extrapolates to a $W$-dependent non-trivial value. These observations indicate a critical phase of the plane defect. As shown in Fig.~\ref{FigS6}(b), at $W \approx 0.73$, $G$ scales as $G \sim L^{-\eta}$ with $\eta=1/4$, which is reminiscent of the anomalous dimension of the Berezinskii-Kosterlitz-Thouless transition. It is natural to ask whether there is a critical phase that features the quasi-long-range order with $0<\eta<1/4$. Figure~\ref{FigS6}(c) demonstrates, with $W=1$, that $\Upsilon L$ is invariant for $K < K_c$. Meanwhile, the exponent $\eta$ is nearly invariant ($\eta \approx 0.16$) [Fig.~\ref{FigS6}(d)]. This result suggests that, as $K \rightarrow 0$, the plane-defect criticality reduces drastically to an essential 2D behavior.

\begin{figure*}
\includegraphics[height=10cm,width=12cm]{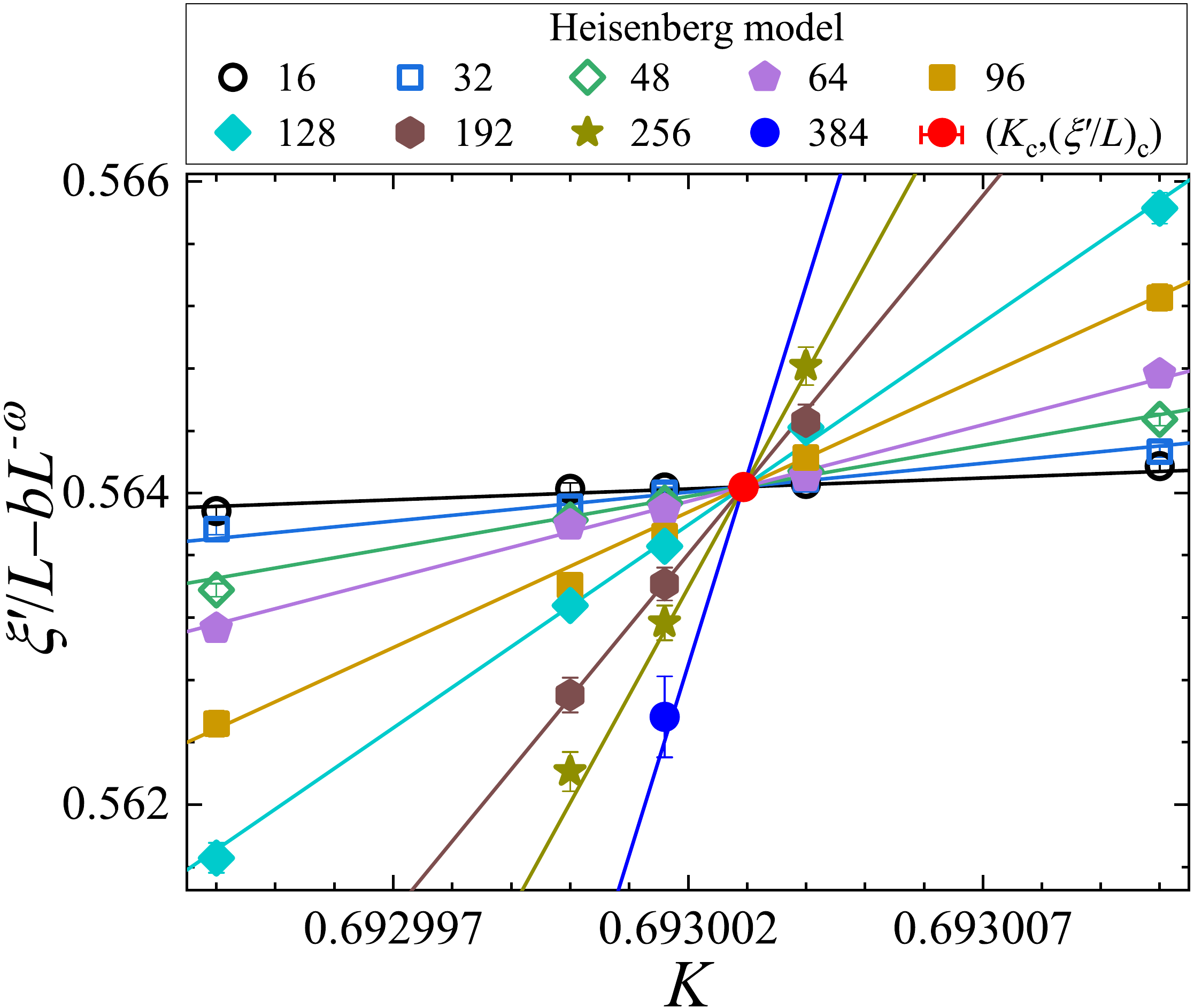}
\caption{The scaled correlation length $\xi'/L$ (with finite-size corrections) versus $K$ for the 3D Heisenberg model. The correction term $bL^{-\omega}$ comes from the preferred fit with $L_{\rm min}=16$ in Table~\ref{tab3dhsb}.}\label{FigS7}
\end{figure*}

\section{Bulk critical point of the 3D Heisenberg model}

To locate the bulk critical point $K_c$ of the Heisenberg model on the simple-cubic lattice, we perform Wolff Monte Carlo simulations with lattice size up to $L=384$. In the simulations, the number of Wolff Monte Carlo steps ranges from $3.8 \times 10^8$ to $1.5 \times 10^{10}$ for $L \leq 64$ and ranges from $6.9 \times 10^8$ to $4.4 \times 10^9$ for $96 \leq L \leq 384$.

For the bulk, the second-moment correlation length $\xi'$ is defined as
\begin{equation}
\xi'=\frac{1}{2{\rm sin}(\pi/L)}\sqrt{\frac{\chi_s}{\chi_1}-1}
\end{equation}
with $\chi_s=\Gamma'(0,0,0)$, $\chi_1=\Gamma'(2\pi/L,0,0)$ and $\Gamma'({\bf k})=\frac{1}{L^3}\langle | \sum_{\bf r}\vec{S}_{\bf r}e^{i{\bf k}\cdot{\bf r}} |^2 \rangle$, where the summation runs over sites on simple-cubic lattices. It is hypothesized that $\xi'/L$ should be scale invariant at $K=K_c$. In the neighborhood of $K_c$, we perform least-squares fits according to
\begin{equation}~\label{Eq1}
\xi'/L=(\xi'/L)_c+a_1(K-K_c)L^{y_t}+bL^{-\omega}
\end{equation}
where $(\xi'/L)_c$ is somewhat universal, whereas $a_1$ and $b$ are non-universal. We find that Monte Carlo data are compatible with the scaling formula. Table~\ref{tab3dhsb} summarizes the results of the fits, for which $(\xi'/L)_c$, $y_t$ and $\omega$ are fixed at the values $(\xi'/L)_c=0.56404$, $y_t=1/0.71164$ and $\omega=0.759$ of 3D Heisenberg universality class~\cite{Hasenbusch2020}. We obtain a set of stable estimates for $K_c$. In particular, we have $K_c=0.69300288(7)$ with $\chi^2/{\rm DoF} \approx 1.0$ and $L_{\rm min}=64$. Figure~\ref{FigS7} shows $\xi'/L$ versus $K$ with finite-size corrections and demonstrates a scale-invariant point.

\section{E-Log critical phase of the 3D plane-defect Heisenberg model}

We explore the plane-defect criticality of 3D Heisenberg model and fix the bulk coupling $K$ at $K_c=0.69300288$. We perform extensive simulations for $W>K_c$ in a broad parameter regime, i.e. $W=2$, $3$ and $7$. The number of Wolff Monte Carlo steps ranges from $3.8 \times 10^8$ to $2.3 \times 10^9$ for $L \leq 64$ and ranges from $3.2 \times 10^8$ to $9.2 \times 10^8$ for $96 \leq L \leq 256$.

We perform fits of $G$ to Eq.~(\ref{seG}) and find that the exponent $\hat{q}$ extrapolates to $\hat{q} \approx 0.63$. In particular, for $W=2$, $3$ and $7$, we obtain $\hat{q}=0.652(2)$, $0.635(1)$ and $0.617(3)$ as well as $\chi^2/{\rm DoF} \approx 1.0$, $1.1$ and $0.3$ with $L_{\rm min}=32$, $16$ and $16$, respectively. For $\chi_s$, we perform fits according to Eq.~(\ref{sechi}). For $W=2$, $3$ and $7$, we obtain $\hat{q}=0.651(4)$, $0.638(2)$ and $0.632(1)$ as well as $\chi^2/{\rm DoF} \approx 0.4$, $0.4$ and $0.4$ with $L_{\rm min}=64$, $32$ and $8$, respectively. Details of fits are given in Tables~\ref{tab_G} and \ref{tab_chi}, for $G$ and $\chi_s$, respectively.

We analyze $\Upsilon$ by performing fits to Eq.~(\ref{seUpsilon}) with $\omega=0.759$. For $W=2$, $3$ and $7$, we obtain $\alpha=0.32(1)$, $0.332(3)$ and $0.340(1)$ as well as $\chi^2/{\rm DoF} \approx 1.0$, $0.6$ and $0.4$ with $L_{\rm min}=96$, $48$ and $32$, respectively. Details of fits are summarized in Table~\ref{tab_Upsilon}.

By comparing the results from the preferred fits in Tables~\ref{tab_G}, \ref{tab_chi} and \ref{tab_Upsilon}, it is inferred that $\hat{q}$ and $\alpha$ are universal. The final estimates of $\hat{q}$ and $\alpha$ are $\hat{q}=0.63(3)$ and $\alpha=0.33(2)$.

\section{E-Log critical phase of the 3D plane-defect O(6) vector model}

We search for the E-Log critical phase of the 3D plane-defect O(6) vector model. We perform simulations at $K=K_c=1.428653$ and $W=3$. The number of Wolff Monte Carlo steps ranges from $1.9 \times 10^8$ to $1.2 \times 10^9$ for $L \leq 64$ and ranges from $2.5 \times 10^8$ to $4.6 \times 10^8$ for $96 \leq L \leq 128$.

For $G$ and $\chi_s$, we perform fits according to Eqs.~(\ref{seG}) and ~(\ref{sechi}), respectively. The results of fits are summarized in Tables~\ref{O6g} and \ref{O6chi}. By comparing the results, we estimate $\hat{q}=2.3(1)$.

We analyze $\Upsilon$ by performing fits to Eq.~(\ref{seUpsilon}), considering the cases with unfixed $\omega$ and without FSS correction, for which consistent results are obtained. The details of fits are listed in Table~\ref{O6H}, and the final estimate of $\alpha$ is $\alpha=0.10(1)$.

\section{Exotic transition and E-Log critical phase of the 4D plane-defect $XY$ model}

We study the plane-defect critical phenomena of the $XY$ model on 4D hypercubic lattices at the bulk critical point $1/K_c=3.314437$. In the Wolff Monte Carlo simulations, the largest system size is $L=96$. The number of Wolff Monte Carlo steps ranges from $1.9 \times 10^8$ to $5.8 \times 10^8$ for $L \leq 40$ and ranges from $4.6 \times 10^7$ to $2.8 \times 10^9$ for $48 \leq L \leq 96$.

For the plane defect, we define the second-moment correlation length
\begin{equation}
\xi=\frac{1}{2{\rm sin}(\pi/L)}\sqrt{\frac{\chi_s}{\chi_1}-1}
\end{equation}
with $\chi_s=\Gamma(0,0)$, $\chi_1=\Gamma(2\pi/L,0)$ and $\Gamma({\bf k})=\frac{1}{L^2}\langle | \sum_{\bf r}\vec{S}_{\bf r}e^{i{\bf k}\cdot{\bf r}} |^2 \rangle$, where the summation runs over sites in the plane defect. For a quantitative analysis of phase transition, we fit the Monte Carlo data of $\xi$ to
\begin{equation}\label{fitxi}
\xi/L=a_0+a_1(W-W_c)L^{y_t}+a_2(W-W_c)^2L^{2y_t}+bL^{-\omega}
\end{equation}
where $W_c$ and $y_t$ denote the transition point and thermal renormalization exponent, respectively. $a_0$, $a_1$, $a_2$ and $b$ are constants. $\omega$ is the exponent for leading finite-size corrections. If $\omega$ is free, we find $W_c=0.411(7)$ and $\omega \approx 1.8(5)$, with $\chi^2/{\rm DoF} \approx 1.6$ and $L_{\rm min}=24$. If $\omega=1$ is fixed, we obtain stable estimates of $W_c$, which include $W_c=0.397(6)$, $0.41(1)$ and $0.41(4)$, with $\chi^2/{\rm DoF} \approx 1.0$, $1.0$ and $0.6$, for $L_{\rm min}=32$, $48$ and $64$, respectively. When $\omega=2$ is fixed, we obtain $W_c=0.413(2)$, $0.411(3)$, $0.416(8)$ and $0.41(2)$, with $\chi^2/{\rm DoF} \approx 1.5$, $1.0$, $1.0$ and $0.5$, for $L_{\rm min}=24$, $32$, $48$ and $64$, respectively. The fits are detailed in Table~\ref{tab_xi4d}. By comparing preferred fits, the final estimates of $W_c$ and $y_t$ are $W_c=0.41(2)$ and $y_t=0.26(3)$, respectively.

We analyze the effects of logarithmic corrections on locating $W_c$. If we consider multiplicative logarithmic corrections with the exponent $1/4$, which arises from the logarithmic corrections of bulk criticality, a scaling formula is given by
\begin{equation}\label{fitxi2}
\xi/[L({\rm ln}L)^{1/4}]=a_0+a_1(W-W_c)L^{y_t}+a_2(W-W_c)^2L^{2y_t}+bL^{-\omega}.
\end{equation}
Stable least-squares fits are achieved, and the results are summarized in Table~\ref{tab_xi4dB}. In particular, we find $W_c=0.424(5)$, $0.43(1)$ and $0.42(3)$, with
$\chi^2/{\rm DoF} \approx 1.2$, $1.2$ and $0.5$, for $L_{\rm min}=32$, $48$ and $64$, respectively. Meanwhile, we estimate $\omega \approx 1.0$. Furthermore, with a reference length $l_0$, one has
\begin{equation}\label{fitxi3}
\xi/[L({\rm ln}(L/l_0))^{1/4}]=a_0+a_1(W-W_c)L^{y_t}+a_2(W-W_c)^2L^{2y_t}+bL^{-\omega}.
\end{equation}
The results of fits are presented in Table~\ref{tab_xi4dC}. We find $W_c=0.41(7)$ and $0.43(4)$, with
$\chi^2/{\rm DoF} \approx 1.3$ and $0.7$, for $L_{\rm min}=48$ and $64$, respectively. By comparing the fits to
Eqs.~(\ref{fitxi}), (\ref{fitxi2}) and (\ref{fitxi3}), we find that the estimates of $W_c$ are close to each other. Hence, the effects of logarithmic corrections on locating $W_c$ are negligible. The existence of a phase transition is further verified.

If $y_t=0$ is assumed, we have a scaling ansatz of $\xi$, which reads
\begin{equation}\label{fitxi4}
\xi/L=a_0+r_1(W-W_c)+r_2(W-W_c)^2+a_1(W-W_c){\rm ln}L+a_2(W-W_c)({\rm ln}L)^2+c(W-W_c)^2{\rm ln}L+b_1L^{-1}+b_2L^{-2},
\end{equation}
where $a_0$, $a_1$, $a_2$, $r_1$, $r_2$, $c$, $b_1$ and $b_2$ are constants. The results of fits are presented in Table~\ref{tab_xi4dd}. With $\chi^2/{\rm DoF} \approx 0.9$, $0.7$ and $0.8$, we find $W_c=0.406(3)$, $0.408(8)$ and $0.42(2)$, with $L_{\rm min}=16$, $24$ and $32$, respectively. Similar results are found with $b_2=0$. These observations are also compatible with the above-mentioned result $W_c=0.41(2)$.

For $W=0.1$, we confirm the scaling behavior $G(L/2) \sim L^{-2}$, which arises from bulk criticality.

We then analyze the cases with $W>W_c$, by simulating at $W=1$, $3$ and $6$. We fit $G$ to Eq.~(\ref{seG}). For $W=1$, $3$ and $6$, we find $\hat{q}=0.093(2)$, $0.084(4)$ and $0.092(3)$ with $\chi^2/{\rm DoF} \approx 0.1$, $0.7$ and $1.7$ as well as $L_{\rm min}=48$, $48$ and $32$, respectively. More results are presented in Table~\ref{tab_G4d}. By comparing preferred fits, the universal value of $\hat{q}$ is estimated as $\hat{q}=0.09(2)$. Furthermore, we find that $\Upsilon$ can be described by the scaling formula
\begin{equation}
\Upsilon L^2=\alpha ({\rm ln}L)^{3/2}+b({\rm ln}L)^{1/2}+c,
\end{equation}
where $\alpha$ is somewhat universal, whereas $b$ and $c$ are non-universal. The fits are presented in Table~\ref{tab_Y4d}. For $W=1$, $3$ and $6$, we find $\alpha=0.93(3)$, $0.97(3)$ and $0.99(4)$ as well as $\chi^2/{\rm DoF} \approx 3.1$, $0.6$ and $0.3$ with $L_{\min}=8$ respectively, which are consistent to each other and indicate E-Log universality. The final estimate of $\alpha$ is $\alpha=0.97(7)$.

\begin{table*}
\caption{Fits of the two-point correlation $G$ to $G=a_0 [{\rm ln}(L/l_0)]^{-\hat{q}}$ for the E-Log critical phase of the 3D plane-defect Villain model at $W=0.5$, $1$, $3$ and $7$. We consider scenarios where $\hat{q}$ is free or held constant at $0.29$.}\label{extG}
\centering
\scalebox{1.0}{
\tabcolsep 0.1in
\begin{tabular}{c|c|c|c|c|c}
\hline
\hline
$W$ &$L_{\rm min}$  &  $\chi^2/{\rm DoF}$  &$a_0$&$l_0$&$\hat{q}$   \\
\hline
\multirow{11}{*}{0.5}&4   &3119.51/5&0.5644(5) &1.535(4)&0.3728(5)\\
&8 & 114.42/4&0.5116(8) &2.22(1)&0.3247(9)\\
&16&  8.48/3&0.494(2) &2.59(4)&0.308(2)\\
&32&  7.97/2&0.490(5) &2.7(1)&0.304(5)\\
&48&  4.72/1&0.47(1) &3.3(4)&0.29(1)\\

&4 & 47488.48/6&0.49014(3) &2.2704(8)&0.29\\
&8 & 2223.24/5&0.47999(5) &2.815(3)&0.29\\
&16 & 125.62/4&0.47685(8) &3.016(5)&0.29\\
&32&  17.53/3&0.4756(1) &3.10(1)&0.29\\
&48&  4.85/2&0.4750(2) &3.15(2)&0.29\\
&64&  4.30/1&0.4748(3) &3.16(3)&0.29\\
\hline
\multirow{11}{*}{1}&4  &345.88/5&1.001(1) &0.418(3)&0.3241(5)\\
&8 & 34.02/4&0.968(2) &0.500(6)&0.3105(9)\\
&16 & 0.34/3&0.945(4) &0.57(1)&0.301(2)\\
&32 & 0.14/2&0.94(1) &0.59(4)&0.299(5)\\
&48& 0.0001/1&0.93(3) &0.6(1)&0.30(1)\\

&4 & 5471.78/6&0.92571(5) &0.6099(4)&0.29\\
&8&  649.07/5&0.92218(7) &0.6463(7)&0.29\\
&16&  42.88/4&0.9198(1) &0.674(1)&0.29\\
&32&  4.16/3&0.9189(2) &0.684(2)&0.29\\
&48&  0.23/2&0.9186(3) &0.689(3)&0.29\\
&64&  0.04/1&0.9184(4) &0.690(5)&0.29\\
\hline
\multirow{11}{*}{3}&4  &9.97/5&1.518(6) &0.0138(5)&0.295(1)\\
&8 & 9.36/4&1.51(1) &0.0143(8)&0.293(2)\\
&16 & 8.47/3&1.49(2) &0.016(2)&0.289(5)\\
&32 & 0.09/2&1.40(3) &0.030(7)&0.268(8)\\
&48 & 0.09/1&1.39(7) &0.03(2)&0.27(2)\\

&4 & 21.33/6&1.4974(1) &0.01546(4)&0.29\\
&8 & 11.58/5&1.4970(2) &0.01557(5)&0.29\\
&16 & 8.49/4&1.4967(3) &0.01567(8)&0.29\\
&32&  6.86/3&1.4964(3) &0.0158(1)&0.29\\
&48&  1.60/2&1.4973(5) &0.0155(2)&0.29\\
&64&  0.81/1&1.4978(8) &0.0153(2)&0.29\\

\hline
\multirow{11}{*}{7}&4 &4.56/5&1.89(3) &0.000025(5)&0.273(4)\\
&8 & 4.50/4&1.88(5) &0.00003(1)&0.272(7)\\
&16 & 4.38/3&1.91(8) &0.00002(1)&0.28(1)\\
&32 & 1.22/2&1.6(1) &0.0002(2)&0.24(2)\\
&48 & 0.94/1&1.8(3) &0.0001(2)&0.26(5)\\

&4 & 18.97/6&2.0108(3) &0.0000109(1)&0.29\\
&8 & 10.24/5&2.0118(5) &0.0000106(1)&0.29\\
&16 & 5.71/4&2.0125(6) &0.0000104(2)&0.29\\
&32&  5.61/3&2.0127(8) &0.0000103(2)&0.29\\
&48&  1.30/2&2.014(1) &0.0000100(3)&0.29\\
&64&  0.31/1&2.015(2) &0.0000096(4)&0.29\\
\hline
\hline
\end{tabular}
}
\end{table*}

\begin{table*}
\caption{Fits of the susceptibility $\chi_s$ to $\chi_s=a_1 L^2 [{\rm ln}(L/l_0)]^{-\hat{q}}$ for the E-Log critical phase of the 3D plane-defect Villain model at $W=0.5$, $1$, $3$ and $7$.}\label{extChi}
\centering
\scalebox{1.0}{
\tabcolsep 0.1in
\begin{tabular}{c|c|c|c|c|c}
\hline
\hline
$W$ &$L_{\rm min}$  &  $\chi^2/{\rm DoF}$  &$a_1$&$l_0$&$\hat{q}$   \\
\hline
\multirow{11}{*}{0.5}&4  &11202.68/5&0.6290(5) &1.354(3)&0.4267(4)\\
&8&  732.74/4&0.5423(7) &2.252(9)&0.3550(6)\\
&16 & 21.14/3&0.507(1) &2.95(3)&0.322(1)\\
&32 & 5.31/2&0.493(4) &3.3(1)&0.309(3)\\
&48 & 3.76/1&0.483(8) &3.7(3)&0.299(8)\\

&4&  215291.53/6&0.49828(2) &2.4563(5)&0.29\\
&8&  15937.48/5&0.48279(4) &3.348(2)&0.29\\
&16&  866.04/4&0.47686(6) &3.780(4)&0.29\\
&32 & 38.33/3&0.4746(1) &3.972(8)&0.29\\
&48 & 5.28/2&0.4738(2) &4.04(1)&0.29\\
&64 & 1.10/1&0.4734(3) &4.08(3)&0.29\\
\hline
\multirow{11}{*}{1} &4 &2091.94/5&1.077(1) &0.356(2)&0.3543(5)\\
&8 & 198.77/4&1.007(2) &0.503(4)&0.3269(7)\\
&16 & 5.21/3&0.964(3) &0.64(1)&0.310(1)\\
&32 & 3.18/2&0.952(9) &0.69(4)&0.305(4)\\
&48 & 1.51/1&0.93(2) &0.8(1)&0.295(8)\\

&4  & 27600.22/6&0.92927(4) &0.7105(4)&0.29\\
&8 & 3487.84/5&0.92361(5) &0.7773(6)&0.29\\
&16&  244.87/4&0.91951(9) &0.832(1)&0.29\\
&32 & 20.79/3&0.9178(1) &0.857(2)&0.29\\
&48 & 1.85/2&0.9172(2) &0.866(3)&0.29\\
&64 & 1.85/1&0.9172(3) &0.866(5)&0.29\\
\hline
\multirow{11}{*}{3} &4  &146.26/5&1.661(6) &0.0076(2)&0.323(1)\\
&8 & 25.48/4&1.581(9) &0.0117(5)&0.308(2)\\
&16 & 9.46/3&1.52(2) &0.016(1)&0.295(3)\\
&32 & 0.47/2&1.44(3) &0.027(5)&0.277(6)\\
&48 & 0.41/1&1.45(6) &0.03(1)&0.28(1)\\

&4 & 1088.78/6&1.5013(1) &0.01733(3)&0.29\\
&8 & 138.99/5&1.4984(1) &0.01834(5)&0.29\\
&16&  12.08/4&1.4967(2) &0.01894(7)&0.29\\
&32 & 4.14/3&1.4963(2) &0.01909(9)&0.29\\
&48 & 0.93/2&1.4967(3) &0.0189(1)&0.29\\
&64 & 0.04/1&1.4972(6) &0.0188(2)&0.29\\
\hline
\multirow{11}{*}{7}&4 &21.84/5&2.19(3) &0.0000039(8)&0.312(4)\\
&8&  9.59/4&2.03(5) &0.000011(3)&0.293(6)\\
&16&  8.91/3&1.99(7) &0.000015(8)&0.29(1)\\
&32&  1.76/2&1.7(1) &0.0002(2)&0.24(2)\\
&48&  0.45/1&1.9(3) &0.00002(6)&0.28(4)\\

&4 & 53.03/6&2.0149(3) &0.00001168(8)&0.29\\
&8&  9.80/5&2.0132(4) &0.0000122(1)&0.29\\
&16 & 9.04/4&2.0130(4) &0.0000123(1)&0.29\\
&32 & 8.42/3&2.0127(6) &0.0000124(2)&0.29\\
&48 & 0.52/2&2.0141(8) &0.0000119(3)&0.29\\
&64 & 0.21/1&2.015(1) &0.0000118(4)&0.29\\
\hline
\hline
\end{tabular}
}
\end{table*}

\begin{table*}
\caption{Fits of the helicity modulus $\Upsilon$ to $\Upsilon L= \alpha ({\rm ln}L) + b + c L^{-\omega}$ for the E-Log critical phase of the 3D plane-defect Villain model at $W=0.5$, $1$, $3$ and $7$. The symbol `-' means that the leading correction term is not included.}\label{extrho}
\centering
\scalebox{1}{
\tabcolsep 0.1in
\begin{tabular}{c|c|c|c|c|c|c}
\hline
\hline
$W$ &$L_{\min}$ &   $\chi^2/{\rm DoF}$  &$\alpha$&$b$ &$c$&$\omega$   \\
\hline
\multirow{14}{*}{0.5}& 16   &805.87/4&0.4780(5)&-0.004(2) &-&-\\
&32 & 55.14/3&0.503(1) &-0.108(4) &-&-\\
&48 & 5.29/2&0.512(2) &-0.148(7) &-&-\\
&64 & 0.16/1&0.516(3) &-0.17(1) &-&-\\

&4 & 134.68/5&0.5362(8)&-0.298(4) &1.254(9)&0.789\\
&8 & 6.96/4&0.549(1) &-0.358(6)&1.47(2)&0.789\\
&16&  0.52/3&0.555(3)&-0.39(1) &1.60(6)&0.789\\
&32&  0.51/2&0.554(7)&-0.38(4) &1.6(2)&0.789\\
&48&  0.35/1&0.55(2)&-0.35(9) &1.3(6)&0.789\\

&4 & 347.10/5&0.5138(7)&-0.177(3) &1.326(9)&1\\
&8 & 31.24/4&0.530(1) &-0.250(5)&1.73(2)&1\\
&16&  0.69/3&0.541(2) &-0.30(1)&2.12(7)&1\\
&32&  0.35/2&0.544(6) &-0.32(3)&2.3(3)&1\\
&48&  0.30/1&0.54(1) &-0.30(7)&2.1(9)&1\\
\hline
\multirow{14}{*}{1}&16 &124.26/4&0.5207(8) &0.633(3) &-&-\\
&32&  13.83/3&0.534(1) &0.578(6) &-&-\\
&48&  0.69/2&0.540(2) &0.55(1) &-&-\\
&64&  0.65/1&0.539(3) &0.55(2) &-&-\\

&4&  8.56/5&0.554(1) &0.462(5)&0.74(1)&0.789\\
&8 & 3.69/4&0.558(2) &0.44(1)&0.80(3)&0.789\\
&16 & 2.12/3&0.562(4)&0.42(2) &0.90(8)&0.789\\
&32 & 2.01/2&0.565(9)&0.41(5) &1.0(3)&0.789\\
&48 & 0.68/1&0.54(2)&0.5(1) &0.1(8)&0.789\\

&4 & 31.92/5&0.541(1) &0.532(4)&0.79(1)&1\\
&8 & 7.88/4&0.548(2) &0.502(8)&0.95(4)&1\\
&16&  2.31/3&0.554(3) &0.47(2)&1.2(1)&1\\
&32&  1.89/2&0.559(7) &0.45(4)&1.5(4)&1\\
&48&  0.68/1&0.54(2) &0.54(9)&0.2(1.3)&1\\

\hline
\multirow{14}{*}{3}&16  &32.79/4&0.548(2) &2.611(7) &-&-\\
&32&  3.59/3&0.557(2)  &2.574(9)&-&-\\
&48&  3.54/2&0.556(4)  &2.58(2)&-&-\\
&64&  1.01/1&0.562(5)  &2.55(2)&-&-\\

&4 & 15.00/5&0.574(2) &2.47(1)&0.68(2)&0.789\\
&8 & 6.32/4&0.582(4)&2.43(2) &0.83(5)&0.789\\
&16 & 6.10/3&0.580(6) &2.45(3)&0.8(1)&0.789\\
&32 & 3.56/2&0.56(1) &2.56(8)&0.1(5)&0.789\\
&48 & 2.33/1&0.59(4) &2.4(2)&1.5(1.4)&0.789\\

&4 & 26.34/5&0.562(2) &2.539(8)&0.73(2)&1\\
&8 & 5.69/4&0.573(3) &2.49(1)&1.00(7)&1\\
&16&  5.67/3&0.574(5) &2.49(3)&1.0(2)&1\\
&32&  3.57/2&0.56(1) &2.57(6)&0.1(7)&1\\
&48&  2.27/1&0.59(3) &2.4(2)&2.4(2.1)&1\\
\hline
\multirow{14}{*}{7}&16  &10.00/4&0.547(3)&6.64(1) &-&- \\
&32&  8.71/3&0.551(4) &6.63(2) &-&-\\
&48&  0.85/2&0.566(7) &6.56(3) &-&-\\
&64&  0.56/1&0.56(1) &6.58(5) &-&-\\

&4 & 6.63/5&0.574(4) &6.51(2)&0.62(5)&0.789\\
&8 & 6.63/4&0.573(7) &6.51(3)&0.6(1)&0.789\\
&16&  6.52/3&0.57(1) &6.53(6)&0.5(3)&0.789\\
&32&  2.09/2&0.63(3) &6.2(2)&2.4(9)&0.789\\
&48&  0.82/1&0.55(7) &6.6(4)&-0.5(2.7)&0.789\\

&4&  7.21/5&0.563(4) &6.56(2)&0.66(5)&1\\
&8 & 6.75/4&0.566(6) &6.55(3)&0.7(1)&1\\
&16&  6.72/3&0.56(1) &6.56(5)&0.7(4)&1\\
&32&  2.00/2&0.61(2) &6.3(1)&3.5(1.3)&1\\
&48&  0.81/1&0.55(6) &6.6(3)&-0.8(4.2)&1\\
\hline
\hline
\end{tabular}
}
\end{table*}

\begin{table*}
\caption{Fits of the two-point correlation $G'$ to $G'=a_0 [{\rm ln}(L/l_0)]^{-\hat{q}}$ for the E-Log critical phase of the 3D plane-defect Villain model at $W=0.5$, $1$, $3$ and $7$.}\label{extGp}
\tabcolsep 0.1in
\begin{tabular}{c|c|c|c|c|c}
\hline
\hline
$W$ &$L_{\rm {min}}$  &  $\chi^2/{\rm DoF}$  &$a_0$&$l_0$&$\hat{q}$   \\
\hline
\multirow{11}{*}{0.5}&4   &26009.53/5&0.6413(4) &1.454(2)&0.4366(4)\\
&8&  596.25/4&0.5356(5) &2.638(8)&0.3471(5)\\
&16&  12.53/3&0.509(1) &3.23(3)&0.322(1)\\
&32 & 7.06/2&0.502(3) &3.43(9)&0.315(3)\\
&48 & 3.72/1&0.490(7) &3.9(3)&0.304(7)\\

&4&  359734.02/6&0.50273(2) &2.6074(4)&0.29\\
&8 & 18915.26/5&0.48562(3) &3.632(2)&0.29\\
&16&  1182.19/4&0.47942(5) &4.107(4)&0.29\\
&32 & 92.33/3&0.47727(8) &4.293(7)&0.29\\
&48 & 7.96/2&0.4762(1) &4.39(1)&0.29\\
&64 & 0.40/1&0.4757(2) &4.44(2)&0.29\\
\hline
\multirow{11}{*}{1}&4  &4332.24/5&1.082(1) &0.415(2)&0.3568(5)\\
&8 & 57.61/4&0.978(2) &0.684(5)&0.3153(7)\\
&16&  2.24/3&0.957(3) &0.77(1)&0.307(1)\\
&32&  1.07/2&0.949(9) &0.81(4)&0.303(4)\\
&48&  0.34/1&0.93(2) &0.9(1)&0.296(8)\\

&4&  36548.61/6&0.93147(4) &0.8081(4)&0.29\\
&8 & 1746.29/5&0.92372(6) &0.9088(7)&0.29\\
&16 & 179.34/4&0.92094(9) &0.950(1)&0.29\\
&32&  14.67/3&0.9193(2) &0.976(2)&0.29\\
&48 & 1.00/2&0.9188(2) &0.985(3)&0.29\\
&64 & 0.77/1&0.9187(3) &0.987(5)&0.29\\
\hline
\multirow{11}{*}{3}&4  &314.26/5&1.611(5) &0.0121(3)&0.314(1)\\
&8 & 4.19/4&1.490(7) &0.0234(9)&0.289(2)\\
&16 & 4.16/3&1.49(1) &0.024(2)&0.288(3)\\
&32 & 3.23/2&1.46(3) &0.028(5)&0.282(7)\\
&48 & 2.08/1&1.53(8) &0.019(8)&0.30(2)\\

&4&  904.58/6&1.49952(9) &0.02125(3)&0.29\\
&8 & 4.90/5&1.4962(1) &0.02261(6)&0.29\\
&16&  4.47/4&1.4963(2) &0.02257(8)&0.29\\
&32 & 4.47/3&1.4963(3) &0.0226(1)&0.29\\
&48 & 2.25/2&1.4967(4) &0.0224(2)&0.29\\
&64 & 2.10/1&1.4969(6) &0.0223(3)&0.29\\
\hline
\multirow{11}{*}{7}&4  &29.95/5&1.95(2) &0.000024(4)&0.281(3)\\
&8&  12.69/4&1.81(4) &0.00007(2)&0.261(5)\\
&16 & 5.42/3&1.96(8) &0.00002(1)&0.28(1)\\
&32 & 1.99/2&1.7(1) &0.0001(1)&0.25(2)\\
&48 & 0.22/1&2.1(5) &0.00001(2)&0.30(6)\\

&4 & 35.54/6&2.0106(3) &0.0000155(1)&0.29\\
&8&  35.45/5&2.0105(4) &0.0000156(2)&0.29\\
&16&  5.87/4&2.0121(5) &0.0000150(2)&0.29\\
&32&  5.82/3&2.0120(7) &0.0000150(3)&0.29\\
&48&  0.28/2&2.014(1) &0.0000143(4)&0.29\\
&64&  0.28/1&2.014(1) &0.0000143(6)&0.29\\
\hline
\hline
\end{tabular}
\end{table*}

\begin{table*}
\caption{Fits of the two-point correlation $G$ to $G=a_0 [{\rm ln}(L/l_0)]^{-\hat{q}}$ for the E-Log critical phase of the 3D plane-defect $XY$ model at $W=1$, $3$ and $7$.}\label{extGXY}
\centering
\scalebox{1.0}{
\tabcolsep 0.1in
\begin{tabular}{c|c|c|c|c|c}
\hline
\hline
$W$ &$L_{\rm min}$  &  $\chi^2/{\rm DoF}$  &$a_0$&$l_0$&$\hat{q}$   \\
\hline
\multirow{9}{*}{1} & 8      & 30.30/4  & 0.7730(7)  & 1.059(5)      & 0.3022(4) \\
& 16          & 10.31/3  & 0.767(1)   & 1.10(1)       & 0.2990(8) \\
& 32           & 0.53/2   & 0.756(4)   & 1.20(3)       & 0.293(2)  \\
& 48           & 0.12/1   & 0.76(1)    & 1.1(1)        & 0.297(7)  \\

& 8            & 935.43/5 & 0.75322(4) & 1.1966(7)     & 0.29      \\
& 16           & 137.58/4 & 0.75200(6) & 1.222(1)      & 0.29      \\
& 32           & 1.95/3   & 0.75114(9) & 1.240(2)      & 0.29      \\
& 48           & 1.10/2   & 0.7510(1)  & 1.242(3)      & 0.29      \\
& 64           & 0.53/1   & 0.7509(3)  & 1.247(7)      & 0.29      \\

\hline
\multirow{9}{*}{3} & 8        & 6.19/4   & 1.429(4)   & 0.0264(6)     & 0.2842(9) \\
& 16    &        6.19/3   & 1.429(7)   & 0.027(1)      & 0.284(2)  \\
& 32    &        1.79/2   & 1.39(2)    & 0.033(4)      & 0.276(4)  \\
& 48    &        0.03/1   & 1.33(5)    & 0.05(2)       & 0.26(1)   \\

& 8     &        43.42/5  & 1.45397(8) & 0.02298(3)    & 0.29      \\
& 16    &        17.98/4  & 1.4544(1)  & 0.02281(5)    & 0.29      \\
& 32    &        12.23/3  & 1.4548(2)  & 0.02266(8)    & 0.29      \\
& 48    &        5.41/2   & 1.4553(3)  & 0.0224(1)     & 0.29      \\
& 64    &        1.38/1   & 1.4564(6)  & 0.0219(3)     & 0.29      \\

\hline
\multirow{9}{*}{7} & 8    & 4.47/4   & 1.88(2)    & 0.000034(4)   & 0.274(2)  \\
& 16    &        2.97/3   & 1.90(3)    & 0.000028(6)   & 0.278(4)  \\
& 32    &        2.23/2   & 1.96(8)    & 0.00002(1)    & 0.29(1)   \\
& 48    &        1.89/1   & 2.1(3)     & 0.00001(1)    & 0.30(4)   \\

& 8     &        52.38/5  & 1.9919(1)  & 0.00001530(6) & 0.29      \\
& 16    &        12.04/4  & 1.9928(2)  & 0.00001495(8) & 0.29      \\
& 32    &        2.38/3   & 1.9936(3)  & 0.0000146(1)  & 0.29      \\
& 48    &        2.07/2   & 1.9938(5)  & 0.0000146(2)  & 0.29      \\
& 64    &        1.09/1   & 1.993(1)   & 0.0000149(4)  & 0.29      \\

\hline
\hline
\end{tabular}
}
\end{table*}

\begin{table*}
\caption{Fits of the helicity modulus $\Upsilon$ to $\Upsilon L= \alpha ({\rm ln}L) + b + c L^{-\omega}$ for the E-Log critical phase of the 3D plane-defect $XY$ model at $W=1$, $3$ and $7$.}\label{extUxy}
\centering
\scalebox{1.0}{
\tabcolsep 0.1in
\begin{tabular}{c|c|c|c|c|c|c}
\hline
\hline
$W$ &$L_{\min}$  &   $\chi^2/{\rm DoF}$  &$\alpha$&$b$ &$c$&$\omega$   \\
\hline
\multirow{13}{*}{1} & 8      & 1202.17/5 & 0.4597(5) & 0.510(1) & -          & -     \\
& 16    &        36.92/4   & 0.498(1)  & 0.393(4) & -          & -     \\
& 32    &        4.22/3    & 0.520(4)  & 0.31(1)  & -          & -     \\
& 48    &        2.55/2    & 0.531(10) & 0.26(4)  & -          & -     \\
& 64    &        0.02/1    & 0.56(2)   & 0.14(9)  & -          & -     \\

& 8     &        1.83/4    & 0.560(3)  & 0.11(1)  & 1.01(3)    & 0.789 \\
& 16    &        1.81/3    & 0.56(1)   & 0.10(5)  & 1.0(2)     & 0.789 \\
& 32    &        1.72/2    & 0.57(3)   & 0.1(2)   & 1.3(8)     & 0.789 \\
& 48    &        0.31/1    & 0.7(1)    & -0.6(6)  & 5.7(38)    & 0.789 \\

& 8     &        2.69/4    & 0.543(2)  & 0.197(9) & 1.14(3)    & 1     \\
& 16    &        1.94/3    & 0.550(9)  & 0.16(4)  & 1.3(2)     & 1     \\
& 32    &        1.76/2    & 0.56(3)   & 0.1(1)   & 1.8(12)    & 1     \\
& 48    &        0.28/1    & 0.66(9)   & -0.4(5)  & 8.7(58)    & 1     \\

\hline
\multirow{13}{*}{3} & 8       & 585.80/5  & 0.5086(5) & 2.533(1) & -          & -     \\
& 16    &        18.72/4   & 0.534(1)  & 2.454(3) & -          & -     \\
& 32    &        8.07/3    & 0.546(4)  & 2.41(1)  & -          & -     \\
& 48    &        5.22/2    & 0.562(10) & 2.35(4)  & -          & -     \\
& 64    &        5.11/1    & 0.56(2)   & 2.37(9)  & -          & -     \\

& 8     &        6.19/4    & 0.575(3)  & 2.27(1)  & 0.66(3)    & 0.789 \\
& 16    &        6.08/3    & 0.57(1)   & 2.28(5)  & 0.6(2)     & 0.789 \\
& 32    &        5.14/2    & 0.60(3)   & 2.1(2)   & 1.4(8)     & 0.789 \\
& 48    &        5.13/1    & 0.6(1)    & 2.2(6)   & 1.1(38)    & 0.789 \\

& 8     &        6.27/4    & 0.564(2)  & 2.326(9) & 0.74(3)    & 1     \\
& 16    &        6.25/3    & 0.565(9)  & 2.32(4)  & 0.8(2)     & 1     \\
& 32    &        5.16/2    & 0.59(3)   & 2.2(1)   & 1.9(11)    & 1     \\
& 48    &        5.15/1    & 0.58(9)   & 2.2(5)   & 1.5(57)    & 1     \\

\hline
\multirow{13}{*}{7} & 8        & 557.53/5  & 0.5207(5) & 6.530(1) & -          & -     \\
& 16    &        13.86/4   & 0.546(1)  & 6.453(3) & -          & -     \\
& 32    &        8.58/3    & 0.554(4)  & 6.42(1)  & -          & -     \\
& 48    &        1.31/2    & 0.579(10) & 6.32(4)  & -          & -     \\
& 64    &        1.28/1    & 0.58(2)   & 6.34(9)  & -          & -     \\

& 8     &        7.18/4    & 0.585(3)  & 6.27(1)  & 0.64(3)    & 0.789 \\
& 16    &        6.11/3    & 0.57(1)   & 6.32(5)  & 0.5(2)     & 0.789 \\
& 32    &        2.15/2    & 0.64(3)   & 6.0(2)   & 2.0(8)     & 0.789 \\
& 48    &        1.09/1    & 0.5(1)    & 6.6(6)   & -1.7(37)   & 0.789 \\

& 8     &        6.69/4    & 0.575(2)  & 6.328(9) & 0.72(3)    & 1     \\
& 16    &        6.34/3    & 0.570(9)  & 6.35(4)  & 0.6(2)     & 1     \\
& 32    &        2.07/2    & 0.62(3)   & 6.1(1)   & 2.9(11)    & 1     \\
& 48    &        1.11/1    & 0.54(9)   & 6.5(5)   & -2.5(57)   & 1     \\

\hline
\hline
\end{tabular}
}
\end{table*}

\begin{table*}
	\caption{Fits of $\xi'$ to $\xi'/L=(\xi'/L)_c+a_1(K-K_c)L^{y_t}+bL^{-\omega}$ for the 3D Heisenberg model [$(\xi'/L)_c=0.56404$, $y_t=1/0.71164$, $\omega=0.759$].}~\label{tab3dhsb}
	\centering
	\scalebox{1.0}{
		\tabcolsep 0.1in
		\begin{tabular}{c|c|c|c|c}
			\hline
			\hline
			$L_{\rm min}$ &  $\chi^2/{\rm DoF}$    & $a_1$       & $K_c$      & $b$        \\
			\hline
			16    & 32.83/34 & 0.285(5) & 0.69300294(6) & -0.0193(1)  \\
			32    & 29.81/29 & 0.284(5) & 0.69300292(7) & -0.0194(2)  \\
			48    & 22.66/24 & 0.287(5) & 0.69300287(7) & -0.0198(3)  \\
			64    & 19.01/19 & 0.285(5) & 0.69300288(7) & -0.0197(4)  \\
			96    & 9.57/14  & 0.285(6) & 0.69300279(9) & -0.021(1)   \\
			128   & 5.47/9   & 0.284(8) & 0.6930028(1)  & -0.021(2)   \\
			192   & 0.60/4   & 0.27(1)  & 0.6930027(2)  & -0.026(7)   \\
			256   & 0.36/1   & 0.27(2)  & 0.6930029(4)  & -0.02(2)    \\
			\hline
			\hline
		\end{tabular}
	}
\end{table*}

\begin{table*}
	\caption{Fits of the two-point correlation $G$ to $G=a_0 [{\rm ln}(L/l_0)]^{-\hat{q}}$ for the E-Log critical phase of the 3D plane-defect Heisenberg model at $W=2$, $3$ and $7$.}~\label{tab_G}
	\centering
	\scalebox{1.0}{
		\tabcolsep 0.1in
		\begin{tabular}{c|c|c|c|c|c}
			\hline
			\hline
			$W$	& $L_{\rm min}$ &  $\chi^2/{\rm DoF}$         & $a_0$       & $l_0$      & $\hat{q}$        \\
			\hline
			\multirow{7}{*}{2} & 8   & 17.98/6 & 1.575(2) & 0.1543(7)    & 0.6582(6) \\
			& 16  & 9.83/5  & 1.566(4) & 0.157(1)     & 0.6561(9) \\
			& 32  & 4.08/4  & 1.548(8) & 0.163(3)     & 0.652(2)  \\
			& 48  & 1.33/3  & 1.53(1)  & 0.171(6)     & 0.647(3)  \\
			& 64  & 0.99/2  & 1.51(3)  & 0.18(1)      & 0.644(6)  \\
			& 96  & 0.68/1  & 1.47(8)  & 0.20(4)      & 0.63(2)   \\
			& 128 & 0.00/0  & 1.3(2)   & 0.3(1)       & 0.60(4)   \\
            \hline
			\multirow{7}{*}{3} & 8   & 14.97/6 & 2.297(5) & 0.0225(2)    & 0.6322(7) \\
			& 16  & 5.45/5  & 2.317(8) & 0.0217(3)    & 0.635(1)  \\
			& 32  & 1.52/4  & 2.29(2)  & 0.0230(7)    & 0.631(2)  \\
			& 48  & 0.03/3  & 2.32(3)  & 0.022(1)     & 0.636(5)  \\
			& 64  & 0.00/2  & 2.31(6)  & 0.022(2)     & 0.634(8)  \\
			& 96  & 0.00/1  & 2.3(2)   & 0.022(7)     & 0.63(2)   \\
			& 128 & 0.00/0  & 2.3(5)   & 0.02(2)      & 0.64(7)   \\
            \hline
			\multirow{7}{*}{7} & 8   & 71.58/6 & 4.15(3)  & 0.0000158(7) & 0.599(2)  \\
			& 16  & 1.54/5  & 4.43(5)  & 0.0000101(7) & 0.617(3)  \\
			& 32  & 0.43/4  & 4.52(10) & 0.000009(1)  & 0.622(6)  \\
			& 48  & 0.22/3  & 4.6(2)   & 0.000008(2)  & 0.63(1)   \\
			& 64  & 0.18/2  & 4.6(3)   & 0.000007(4)  & 0.63(2)   \\
			& 96  & 0.02/1  & 5.0(11)  & 0.000004(6)  & 0.65(6)   \\
			& 128 & 0.00/0  & 5.4(34)  & 0.00000(1)   & 0.7(2)      \\
			\hline
			\hline
		\end{tabular}
	}
\end{table*}

\begin{table*}
	\caption{Fits of the susceptibility $\chi_s$ to $\chi_s=a_1 L^2 [{\rm ln}(L/l_0)]^{-\hat{q}}$ for the E-Log critical phase of the 3D plane-defect Heisenberg model at $W=2$, $3$ and $7$.}~\label{tab_chi}
	\centering
	\scalebox{1.0}{
		\tabcolsep 0.1in
		\begin{tabular}{c|c|c|c|c|c}
			\hline
			\hline
			$W$	& $L_{\rm min}$ &  $\chi^2/{\rm DoF}$         & $a_1$       & $l_0$      & $\hat{q}$        \\
			\hline
			\multirow{7}{*}{2} & 8   & 1305.05/6 & 1.715(2) & 0.1458(5)    & 0.6897(4) \\
			& 16  & 108.87/5  & 1.629(3) & 0.1715(10)   & 0.6711(7) \\
			& 32  & 10.06/4   & 1.575(6) & 0.192(2)     & 0.659(1)  \\
			& 48  & 1.02/3    & 1.55(1)  & 0.203(5)     & 0.653(2)  \\
			& 64  & 0.74/2    & 1.54(2)  & 0.207(9)     & 0.651(4)  \\
			& 96  & 0.20/1    & 1.50(6)  & 0.23(3)      & 0.64(1)   \\
			& 128 & 0.00/0    & 1.6(2)   & 0.19(8)      & 0.66(4)   \\
              \hline
			\multirow{7}{*}{3} & 8   & 308.99/6  & 2.508(5) & 0.0194(1)    & 0.6610(6) \\
			& 16  & 48.24/5   & 2.420(7) & 0.0223(2)    & 0.6494(9) \\
			& 32  & 1.68/4    & 2.34(1)  & 0.0257(6)    & 0.638(2)  \\
			& 48  & 1.05/3    & 2.36(2)  & 0.025(1)     & 0.641(3)  \\
			& 64  & 0.70/2    & 2.33(4)  & 0.026(2)     & 0.638(6)  \\
			& 96  & 0.02/1    & 2.2(1)   & 0.031(7)     & 0.62(2)   \\
			& 128 & 0.00/0    & 2.3(4)   & 0.03(2)      & 0.63(5)   \\
              \hline
			\multirow{7}{*}{7} & 8   & 2.59/6    & 4.69(3)  & 0.0000083(3) & 0.632(1)  \\
			& 16  & 2.14/5    & 4.67(4)  & 0.0000085(5) & 0.630(2)  \\
			& 32  & 0.97/4    & 4.60(8)  & 0.000010(1)  & 0.626(4)  \\
			& 48  & 0.97/3    & 4.6(1)   & 0.000010(2)  & 0.626(8)  \\
			& 64  & 0.85/2    & 4.5(2)   & 0.000011(4)  & 0.62(1)   \\
			& 96  & 0.26/1    & 5.1(9)   & 0.000005(6)  & 0.65(4)   \\
			& 128 & 0.00/0    & 6.4(35)  & 0.000001(4)  & 0.7(1)     \\
			\hline
			\hline
		\end{tabular}
	}
\end{table*}

\begin{table*}
	\caption{Fits of the helicity modulus $\Upsilon$ to $\Upsilon L= \alpha ({\rm ln}L) + b + c L^{-\omega}$ for the E-Log critical phase of the 3D plane-defect Heisenberg model at $W=2$, $3$ and $7$ ($\omega=0.759$).}~\label{tab_Upsilon}
	\centering
	\scalebox{1.0}{
		\tabcolsep 0.1in
		\begin{tabular}{c|c|c|c|c|c}
			\hline
			\hline
			$W$	& $L_{\rm min}$ &  $\chi^2/{\rm DoF}$         & $\alpha$       & $b$      & $c$        \\
			\hline
			\multirow{13}{*}{2}
			& 16  & 101.30/6  & 0.3026(4) & 0.909(1)  & -        \\
			& 32  & 8.00/5    & 0.314(1)  & 0.868(4)  & -        \\
			& 48  & 3.42/4    & 0.319(3)  & 0.85(1)   & -        \\
			& 64  & 1.95/3    & 0.324(5)  & 0.82(2)   & -        \\
			& 96  & 1.95/2    & 0.32(1)   & 0.82(5)   & -        \\
			& 128 & 1.48/1    & 0.34(2)   & 0.77(10)  & -        \\
			& 192 & 0.00/0    & 0.41(7)   & 0.3(4)    & -        \\
			& 8   & 3.59/6    & 0.338(1)  & 0.742(4)  & 0.56(1)  \\
			& 16  & 2.14/5    & 0.335(3)  & 0.76(2)   & 0.50(5)  \\
			& 32  & 2.14/4    & 0.335(9)  & 0.76(4)   & 0.5(2)   \\
			& 48  & 1.96/3    & 0.34(2)   & 0.7(1)    & 0.7(6)   \\
			& 64  & 1.89/2    & 0.33(4)   & 0.8(2)    & 0.4(15)  \\
			& 96  & 1.07/1    & 0.4(1)    & 0.2(6)    & 5.3(56)  \\
              \hline
			\multirow{13}{*}{3}
			& 16  & 71.81/6   & 0.3171(4) & 1.578(1)  & -        \\
			& 32  & 7.31/5    & 0.326(1)  & 1.544(4)  & -        \\
			& 48  & 2.44/4    & 0.332(3)  & 1.52(1)   & -        \\
			& 64  & 1.12/3    & 0.336(5)  & 1.50(2)   & -        \\
			& 96  & 0.76/2    & 0.34(1)   & 1.47(5)   & -        \\
			& 128 & 0.15/1    & 0.35(2)   & 1.41(10)  & -        \\
			& 192 & 0.00/0    & 0.38(7)   & 1.3(4)    & -        \\
			& 8   & 2.81/6    & 0.348(1)  & 1.433(4)  & 0.49(1)  \\
			& 16  & 1.06/5    & 0.344(3)  & 1.45(2)   & 0.42(5)  \\
			& 32  & 0.83/4    & 0.348(9)  & 1.43(4)   & 0.5(2)   \\
			& 48  & 0.54/3    & 0.36(2)   & 1.4(1)    & 0.8(6)   \\
			& 64  & 0.48/2    & 0.36(4)   & 1.3(2)    & 1.1(14)  \\
			& 96  & 0.05/1    & 0.42(10)  & 1.0(6)    & 4.6(55)  \\
              \hline
			\multirow{13}{*}{7}
			& 16  & 57.00/6   & 0.3318(4) & 4.244(1)  & -        \\
			& 32  & 1.90/5    & 0.340(1)  & 4.213(4)  & -        \\
			& 48  & 0.24/4    & 0.343(3)  & 4.20(1)   & -        \\
			& 64  & 0.17/3    & 0.344(5)  & 4.20(2)   & -        \\
			& 96  & 0.16/2    & 0.34(1)   & 4.20(5)   & -        \\
			& 128 & 0.00/1    & 0.35(2)   & 4.17(9)   & -        \\
			& 192 & 0.00/0    & 0.35(6)   & 4.2(3)    & -        \\
			& 8   & 2.38/6    & 0.360(1)  & 4.114(4)  & 0.44(1)  \\
			& 16  & 0.61/5    & 0.356(3)  & 4.13(1)   & 0.38(5)  \\
			& 32  & 0.23/4    & 0.351(8)  & 4.16(4)   & 0.3(2)   \\
			& 48  & 0.18/3    & 0.35(2)   & 4.18(10)  & 0.1(6)   \\
			& 64  & 0.17/2    & 0.34(3)   & 4.2(2)    & -0.0(14) \\
			& 96  & 0.03/1    & 0.38(10)  & 4.0(6)    & 2.0(55)     \\
			\hline
			\hline
		\end{tabular}
	}
\end{table*}

\begin{table*}
	\caption{Fits of the two-point correlation $G$ to $G=a_0 [{\rm ln}(L/l_0)]^{-\hat{q}}$ for the E-Log critical phase of the 3D plane-defect O(6) vector model at $W=3$.}~\label{O6g}
	\centering
	\scalebox{1.0}{
		\tabcolsep 0.1in
		\begin{tabular}{c|c|c|c|c}
			\hline
			\hline
			$L_{\rm min}$ & $\chi^2$/DoF         & $a_0$       & $l_0$      & $\hat{q}$\\
			\hline
			8  & 9.78/4 & 31.8(5)  & 0.0118(2) & 2.278(5) \\
			16 & 1.14/3 & 34.1(10) & 0.0109(4) & 2.300(9) \\
			32 & 1.00/2 & 35.0(25) & 0.0106(9) & 2.31(2)  \\
			48 & 0.94/1 & 33.4(63) & 0.011(3)  & 2.29(6)   \\
			\hline
			\hline
		\end{tabular}
	}
\end{table*}

\begin{table*}
	\caption{Fits of the susceptibility $\chi_s$ to $\chi_s=a_1 L^2 [{\rm ln}(L/l_0)]^{-\hat{q}}$ for the E-Log critical phase of the 3D plane-defect O(6) vector model at $W=3$.}~\label{O6chi}
	\centering
	\scalebox{1.0}{
		\tabcolsep 0.1in
		\begin{tabular}{c|c|c|c|c}
			\hline
			\hline
			$L_{\rm min}$ & $\chi^2$/DoF         & $a_1$       & $l_0$      & $\hat{q}$\\
			\hline
			8  & 317.64/4 & 68.3(9)  & 0.00619(9) & 2.522(4) \\
			16 & 14.07/3  & 49.0(11) & 0.0090(2)  & 2.417(7) \\
			32 & 0.49/2   & 40.9(21) & 0.0111(7)  & 2.36(2)  \\
			48 & 0.47/1   & 40.1(55) & 0.011(2)   & 2.35(4)  \\
			\hline
			\hline
		\end{tabular}
	}
\end{table*}

\begin{table*}
	\caption{Fits of the helicity modulus $\Upsilon$ to $\Upsilon L= \alpha ({\rm ln}L) + b + c L^{-\omega}$ for the E-Log critical phase of the 3D plane-defect O(6) vector model at $W=3$.}~\label{O6H}
	\centering
	\scalebox{1.0}{
		\tabcolsep 0.1in
		\begin{tabular}{c|c|c|c|c|c|c}
			\hline
			\hline
			$L_{\rm min}$ & $\chi^2$/DoF         & $\alpha$       & $b$      & $c$       & $\omega$ \\
			\hline
			8  & 2.75/3    & 0.108(3)  & 0.70(1)   & 0.5(2) & 1.4(3) \\
			16 & 29.61/4   & 0.0988(4) & 0.740(1)  & -     & -     \\
			32 & 1.57/3    & 0.104(1)  & 0.721(4)  & -     & -     \\
			48 & 1.43/2    & 0.103(2)  & 0.724(9)  & -     & -     \\
			64 & 0.88/1    & 0.106(5)  & 0.71(2)   & -     & -     \\
			\hline
			\hline
		\end{tabular}
	}
\end{table*}

\begin{table*}
	\caption{Fits of the correlation length $\xi$ to $\xi/L=a_0+a_1(W-W_c)L^{y_t}+a_2(W-W_c)^2L^{2y_t}+bL^{-\omega}$ for the 4D plane-defect $XY$ model.}~\label{tab_xi4d}
	\centering
	\scalebox{1.0}{
		\tabcolsep 0.1in
	\begin{tabular}{c|c|c|c|c|c|c|c|c}
			\hline
			\hline
			$L_{\rm min}$ &  $\chi^2/{\rm DoF}$       & $a_0$       & $a_1$      & $W_c$   & $y_t$   & $a_2$  & $b$  & $\omega$\\
			\hline
			16 & 88.18/14   & 0.210(7)  & 0.44(2)  & 0.402(6)  & 0.204(9)  & 1.18(8) & 1.3(4)    & 1.5(1) \\
			24 & 17.29/11   & 0.220(10) & 0.41(1)  & 0.411(7)  & 0.25(2)   & 0.9(1)  & 3.0(39)   & 1.8(5) \\
			16 & 105.49/15  & 0.177(2)  & 0.34(1)  & 0.371(3)  & 0.168(5)  & 1.47(8) & 0.559(8)  & 1      \\
			24 & 21.81/12   & 0.194(4)  & 0.37(1)  & 0.391(4)  & 0.212(8)  & 1.14(8) & 0.48(2)   & 1      \\
			32 & 9.44/9     & 0.201(7)  & 0.37(2)  & 0.397(6)  & 0.23(1)   & 1.1(1)  & 0.43(6)   & 1      \\
			48 & 5.85/6     & 0.22(2)   & 0.36(3)  & 0.41(1)   & 0.27(4)   & 0.8(3)  & 0.3(3)    & 1      \\
			64 & 1.65/3     & 0.21(6)   & 0.30(5)  & 0.41(4)   & 0.3(1)    & 0.5(5)  & 0.4(10)   & 1      \\
			16 & 102.06/15  & 0.228(1)  & 0.472(5) & 0.4167(9) & 0.227(4)  & 1.01(4) & 4.20(6)   & 2      \\
			24 & 17.48/12   & 0.223(2)  & 0.417(7) & 0.413(2)  & 0.250(7)  & 0.87(5) & 5.1(3)    & 2      \\
			32 & 8.63/9     & 0.221(4)  & 0.39(1)  & 0.411(3)  & 0.26(1)   & 0.88(9) & 5.9(8)    & 2      \\
			48 & 6.20/6     & 0.23(1)   & 0.37(2)  & 0.416(8)  & 0.29(3)   & 0.7(2)  & 4.0(50)   & 2      \\
			64 & 1.59/3     & 0.22(3)   & 0.30(5)  & 0.41(2)   & 0.32(8)   & 0.5(4)  & 11.2(252) & 2      \\
			32 & 58.74/10   & 0.2452(8) & 0.40(1)  & 0.4293(6) & 0.305(10) & 0.60(5) & -         & -      \\
			48 & 6.85/7     & 0.235(2)  & 0.37(3)  & 0.422(1)  & 0.30(2)   & 0.6(1)  & -         & -      \\
			64 & 1.78/4     & 0.235(4)  & 0.29(5)  & 0.422(3)  & 0.36(4)   & 0.4(1)  & -         & -      \\
			80 & 0.04/1     & 0.22(2)   & 0.5(2)   & 0.41(1)   & 0.2(1)    & 1.2(12) & -         & -      \\
			\hline
			\hline
		\end{tabular}
	}
\end{table*}

\begin{table*}
	\caption{Fits of the correlation length $\xi$ to $\xi/[L({\rm ln}L)^{1/4}]=a_0+a_1(W-W_c)L^{y_t}+a_2(W-W_c)^2L^{2y_t}+bL^{-\omega}$ for the 4D plane-defect $XY$ model.}~\label{tab_xi4dB}
	\centering
	\scalebox{1.0}{
		\tabcolsep 0.1in
		\begin{tabular}{c|c|c|c|c|c|c|c|c}
			\hline
			\hline
			$L_{\rm min}$ & $\chi^2/{\rm DoF}$      & $a_0$       & $a_1$      & $W_c$   & $y_t$   & $a_2$  & $b$  & $\omega$\\
			\hline
			16 & 124.55/14  & 0.166(9) & 0.46(1)  & 0.430(8)  & 0.18(1)  & 1.04(9)  & 0.8(1)    & 1.16(8) \\
			24 & 21.68/11   & 0.17(1)  & 0.398(8) & 0.433(10) & 0.23(2)  & 0.8(1)   & 0.9(4)    & 1.2(2)  \\
			32 & 10.48/8    & 0.15(3)  & 0.36(2)  & 0.42(2)   & 0.22(5)  & 0.8(3)   & 0.5(5)    & 0.9(4)  \\
			16 & 130.26/15  & 0.146(2) & 0.427(7) & 0.411(2)  & 0.152(5) & 1.23(5)  & 0.635(6)  & 1       \\
			24 & 22.87/12   & 0.156(3) & 0.391(8) & 0.422(3)  & 0.204(8) & 0.88(6)  & 0.59(2)   & 1       \\
			32 & 10.49/9    & 0.158(5) & 0.36(1)  & 0.424(5)  & 0.23(1)  & 0.78(10) & 0.57(5)   & 1       \\
			48 & 7.03/6     & 0.16(2)  & 0.32(3)  & 0.43(1)   & 0.27(4)  & 0.6(2)   & 0.5(2)    & 1       \\
			64 & 1.50/3     & 0.16(4)  & 0.26(6)  & 0.42(3)   & 0.3(1)   & 0.4(4)   & 0.6(8)    & 1       \\
			\hline
			\hline
		\end{tabular}
	}
\end{table*}

\begin{table*}
	\caption{Fits of the correlation length $\xi$ to $\xi/[L({\rm ln}(L/l_0))^{1/4}]=a_0+a_1(W-W_c)L^{y_t}+a_2(W-W_c)^2L^{2y_t}+bL^{-\omega}$ for the 4D plane-defect $XY$ model.}~\label{tab_xi4dC}
	\centering
	\scalebox{1.0}{
		\tabcolsep 0.1in
		\begin{tabular}{c|c|c|c|c|c|c|c|c|c}
			\hline
			\hline
			$L_{\rm min}$ & $\chi^2/{\rm DoF}$    & $l_0$    & $a_0$       & $a_1$      & $W_c$   & $y_t$   & $a_2$  & $b$  & $\omega$\\
			\hline
			48 & 6.54/5    & 16.1(66)  	  & 0.17(8)  & 0.7(2)    & 0.41(7)  & 0.1(1)   & 2.6(30) & 2.6(14) & 1 \\
			64 & 1.39/2    & 11.8(442) 	  & 0.19(7)  & 0.4(8)    & 0.43(4)  & 0.2(4)   & 0.8(28) & 1.9(57) & 1   \\
			
			\hline
			\hline
		\end{tabular}
	}
\end{table*}

\begin{table*}
	\caption{Fits of the correlation length $\xi$ to $\xi/L=a_0+r_1(W-W_c)+r_2(W-W_c)^2+a_1(W-W_c){\rm ln}L+a_2(W-W_c)({\rm ln}L)^2+c(W-W_c)^2{\rm ln}L+b_1L^{-1}+b_2L^{-2}$ for the 4D plane-defect $XY$ model.}~\label{tab_xi4dd}
	\centering
	\scalebox{1.0}{
		\tabcolsep 0.1in
		\begin{tabular}{c|c|c|c|c|c|c|c|c|c|c}
			\hline
			\hline
			$L_{\rm min}$ & $\chi^2/{\rm DoF}$    & $a_0$    & $r_1$       & $W_c$      & $r_2$   & $a_1$   & $a_2$  & $c$  & $b_1$ & $b_2$\\
			\hline
			16     & 10.53/12 & 0.212(4) & 1.6(1)  & 0.406(3) & -4.8(10) & -0.56(6) & 0.106(8) & 2.9(3)  & 0.24(6)  & 2.0(5)    \\
			24     & 6.14/9   & 0.22(1)  & 1.6(3)  & 0.408(8) & -7.4(16) & -0.5(1)  & 0.10(2)  & 3.6(5)  & 0.2(2)   & 2.8(24)   \\
			32     & 4.96/6   & 0.24(4)  & 1.7(6)  & 0.42(2)  & -7.1(29) & -0.6(3)  & 0.13(4)  & 3.5(8)  & -0.6(9)  & 14.5(143) \\
			16     & 24.86/13 & 0.197(2) & 1.7(1)  & 0.394(2) & -4.8(10) & -0.63(5) & 0.106(8) & 2.9(3)  & 0.450(7) & -         \\
			24     & 7.48/10  & 0.203(4) & 1.7(2)  & 0.399(4) & -7.5(16) & -0.6(1)  & 0.10(2)  & 3.6(5)  & 0.40(2)  & -         \\
			32     & 6.04/7   & 0.210(6) & 2.0(6)  & 0.405(5) & -7.2(29) & -0.7(3)  & 0.12(4)  & 3.5(8)  & 0.35(5)  & -         \\
			48     & 3.60/4   & 0.23(1)  & 3.1(24) & 0.417(8) & -1.2(69) & -1.3(12) & 0.2(1)   & 2.1(17) & 0.1(2)   & -			\\
			
			\hline
			\hline
		\end{tabular}
	}
\end{table*}

\newpage

\begin{table*}
	\caption{Fits of the two-point correlation $G$ to $G=a_0 [{\rm ln}(L/l_0)]^{-\hat{q}}$ for the E-Log critical phase of the 4D plane-defect $XY$ model at $W=1$, $3$ and $6$.}~\label{tab_G4d}
	\centering
	\scalebox{1.0}{
		\tabcolsep 0.1in
		\begin{tabular}{c|c|c|c|c|c}
			\hline
			\hline
			$W$	& $L_{\rm min}$ &  $\chi^2/{\rm DoF}$         & $a_0$       & $l_0$      &  $\hat{q}$     \\
			\hline
			\multirow{6}{*}{1}
			& 16    & 728.93/5  & 0.5722(3) & 5.51(3)  & 0.1218(3) \\
			& 24    & 109.09/4  & 0.5602(5) & 6.84(6)  & 0.1095(5) \\
			& 32    & 18.67/3   & 0.5522(8) & 8.0(1)   & 0.1012(9) \\
			& 40    & 7.40/2    & 0.549(1)  & 8.7(2)   & 0.097(1)  \\
			& 48    & 0.13/1    & 0.545(2)  & 9.5(4)   & 0.093(2)  \\
			& 64    & 0.00/0    & 0.542(7)  & 10.1(16) & 0.091(7)  \\
           \hline
			\multirow{6}{*}{3}
			& 16    & 108.35/5  & 0.933(1)  & 1.31(2)  & 0.1096(6) \\
			& 24    & 25.43/4   & 0.916(2)  & 1.73(6)  & 0.101(1)  \\
			& 32    & 11.36/3   & 0.903(4)  & 2.2(1)   & 0.095(2)  \\
			& 40    & 2.82/2    & 0.890(5)  & 2.7(2)   & 0.088(3)  \\
			& 48    & 0.74/1    & 0.882(7)  & 3.2(4)   & 0.084(4)  \\
			& 64    & 0.00/0    & 0.86(2)   & 4.5(17)  & 0.075(10) \\
                 \hline
			\multirow{6}{*}{6}
			& 16    & 36.87/5   & 1.074(3)  & 0.20(1)  & 0.106(1)  \\
			& 24    & 7.74/4    & 1.049(5)  & 0.31(3)  & 0.097(2)  \\
			& 32    & 5.07/3    & 1.037(8)  & 0.39(7)  & 0.092(3)  \\
			& 40    & 3.87/2    & 1.02(1)   & 0.5(2)   & 0.087(5)  \\
			& 48    & 1.70/1    & 1.01(2)   & 0.7(2)   & 0.081(6)  \\
			& 64    & 0.00/0    & 1.1(1)    & 0.1(2)   & 0.12(4)    \\
			\hline
			\hline
		\end{tabular}
	}
\end{table*}

\newpage
\begin{table*}
	\caption{Fits of the helicity modulus $\Upsilon$ to $\Upsilon L^2=\alpha({\rm ln}L)^{3/2}+b({\rm ln}L)^{1/2}+c$ for the E-Log critical phase of the 4D plane-defect $XY$ model at $W=1$, $3$ and $6$.}~\label{tab_Y4d}
	\centering
	\scalebox{1.0}{
		\tabcolsep 0.1in
		\begin{tabular}{c|c|c|c|c|c}
			\hline
			\hline
			$W$	& $L_{\rm min}$ & $\chi^2/{\rm DoF}$        & $\alpha$       & $b$      & $c$        \\
			\hline
			\multirow{8}{*}{1}  & 8  & 18.73/6 & 0.93(3)  & -3.2(2)     & 3.8(3)     \\
			& 16 & 17.96/5 & 1.0(1)   & -4.1(11)    & 4.9(13)    \\
			& 8  & 20.37/7 & 0.97 & -3.50(2) & 4.16(2)   \\
			& 16 & 18.19/6 & 0.97 & -3.59(7) & 4.3(1)    \\
			& 24 & 18.11/5 & 0.97 & -3.6(2)  & 4.4(3)    \\
			& 32 & 15.36/4 & 0.97 & -3.1(3)  & 3.4(7)    \\
			& 40 & 14.73/3 & 0.97 & -2.7(6)  & 2.7(12)   \\
			& 48 & 12.12/2 & 0.97 & -1.7(9)  & 0.5(18)   \\
           \hline
			\multirow{8}{*}{3} & 8  & 3.75/6  & 0.97(3)  & -3.0(3)     & 5.6(3)     \\
			& 16 & 2.23/5  & 1.1(1)   & -4.4(11)    & 7.3(13)    \\
			& 8  & 3.76/7  & 0.97 & -3.03(1) & 5.68(2)   \\
			& 16 & 3.55/6  & 0.97 & -3.06(7) & 5.7(1)    \\
			& 24 & 2.63/5  & 0.97 & -2.9(2)  & 5.5(3)    \\
			& 32 & 1.99/4  & 0.97 & -2.6(4)  & 4.9(7)    \\
			& 40 & 1.51/3  & 0.97 & -2.3(7)  & 4.2(13)   \\
			& 48 & 0.97/2  & 0.97 & -2.8(10) & 5.4(21)   \\
                     \hline
			\multirow{8}{*}{6} & 8  & 1.98/6  & 0.99(4)  & -3.1(3)     & 8.8(3)     \\
			& 16 & 0.34/5  & 1.1(1)   & -4.6(12)    & 10.5(14)   \\
			& 8  & 2.43/7  & 0.97 & -2.95(2) & 8.60(2)   \\
			& 16 & 2.35/6  & 0.97 & -2.93(7) & 8.6(1)    \\
			& 24 & 0.61/5  & 0.97 & -2.7(2)  & 8.1(3)    \\
			& 32 & 0.53/4  & 0.97 & -2.6(4)  & 8.0(7)    \\
			& 40 & 0.29/3  & 0.97 & -2.3(8)  & 7.3(15)   \\
			& 48 & 0.17/2  & 0.97 & -2.0(10) & 6.9(20)   \\
			\hline
			\hline
		\end{tabular}
	}
\end{table*}